\documentclass[letterpaper]{article}

\usepackage[T1]{fontenc}

\usepackage{geometry}
\usepackage{setspace}
\usepackage{tabularx}

\usepackage[style = chem-acs]{biblatex}
\usepackage{graphicx}
\usepackage{float}
\newfloat{scheme}{htbp}{los}
\floatname{scheme}{Scheme}
\floatname{chart}{Chart}
\newfloat{graph}{htbp}{loh}

\usepackage{chemformula} 
\usepackage[version = 4]{mhchem} 

\usepackage{authblk}

\author[1*$\dagger$]{Fabio Bersano}
\author[1$\dagger$]{Cyrille Masserey} 
\author[1$\dagger$]{Vanessa Conti}
\author[1]{Andrea Iaconeta}
\author[1]{Niccolò Martinolli}
\author[1]{Ehsan Ansari}
\author[1]{Anna Varini}
\author[1]{Igor Stolichnov}
\author[1*]{Adrian Mihai Ionescu}

\affil[1]{Institute of Electrical and Micro Engineering, École Polytechnique Fédérale de Lausanne (EPFL), Lausanne, 1015, Switzerland}

\title{Monolithically Integrated VO$_2$ Mott Oscillators for Energy-Efficient Spiking Neurons}
\date{$\dagger$ These authors contributed equally to this work.\\
*Email: fabio.bersano@epfl.ch, adrian.ionescu@epfl.ch}

\begin{document}
\begin{refsection}

\maketitle

\begin{abstract}
  Brain-inspired non-Boolean computing and sensing enable energy-efficient, error-tolerant, and highly parallel information processing, yet their practical deployment remains limited by the lack of compact spiking hardware compatible with large-scale integration. Mott phase-transition materials provide a promising device-level route, as their abrupt insulator-to-metal transitions enable neuron-like thresholding and oscillatory dynamics. Among these, vanadium dioxide (VO$_2$) stands out for its near-room-temperature transition, fast switching, and scalability. However, existing VO$_2$-based neuristor implementations rely on discrete components, constraining integration density and system-level applicability. Here, we report monolithic back-end-of-the-line (BEOL) integration of one-transistor-one-VO$_2$-memristor (1T-1MR) spiking neurons directly on a functional CMOS-compatible platform. VO$_2$ nanosheet devices are fabricated by pulsed-laser deposition atop dielectrically isolated silicon-on-insulator (SOI) p-type junctionless field-effect transistors (JLFETs) at temperatures below 430~$^\circ$C. This scalable architecture exhibits gate-tunable oscillations spanning 40-410~kHz in 60~nm-thin VO$_2$ devices with an active area of 6~\textmu m$^2$, achieving memristor energy and power consumption as low as 18~pJ per spike and 8~\textmu W at room temperature, with the potential to scale below sub-3~\textmu W operation. We uncover a non-monotonic dependence of oscillation frequency on bias current and temperature and analyze bias-dependent stochastic firing dynamics, revealing the nonlinear physics of integrated VO$_2$ thin-film memristor systems. Finally, we demonstrate voltage-controlled oscillator functionality and on-chip active resistive coupling of two nano-oscillators mediated by a JLFET. These results establish a viable pathway toward dense, energy-efficient, and monolithically integrated Mott-based neuromorphic hardware, bridging emerging phase-transition devices with CMOS technology for future computing and spiking sensing systems.
\end{abstract}


\newpage
\section{Introduction}

Traditional computing systems based on the von Neumann (VN) architecture have long dominated digital hardware design due to their instruction-level programmability, sequential control flow, and compatibility with deeply scaled CMOS implementations supported by hierarchical memory integration~\cite{waldrop2016chips, williams2017s}. This paradigm provides a flexible and customizable framework for general-purpose data processing, enabling broad software-hardware co-optimization across application domains. Nevertheless, these systems fundamentally suffer from energy and latency overheads associated with extensive data movement between physically separated memory and processing units, a bottleneck that becomes particularly severe in data-centric and edge-processing machine learning (ML) and artificial intelligence (AI) workloads~\cite{williams2017s,kendall2020building,christensen20222022}.
This limitation has stimulated the development of low-power AI accelerators that enable parallel processing of data streams rather than sequential computation. Such approaches mainly rely on artificial neural network (ANN), deep neural network (DNN), and spiking neural network (SNN) algorithms implemented through neuro-inspired hardware architectures~\cite{sze2017efficient, zhang2020neuro, kendall2020building}. In these systems, computation is distributed across electrical networks of artificial neurons interconnected by programmable synapses that locally store and weight information, enabling the co-location of memory and processing~\cite{chicca2014neuromorphic,burr2017neuromorphic,zhang2020neuro,kendall2020building}. While ANN and DNN accelerators typically encode neuronal states as digital activation values, SNNs represent information through the timing of analog discrete electrical spikes. This temporal and event-driven encoding enables sparse communication for fast inference, low power consumption, and efficient interaction with biomimetic spiking sensory signals~\cite{kim2018bioinspired,zhang2020artificial,markiewicz2024spiking,zhou2023computational,subbulakshmi2021biomimetic}. These advantages have motivated the exploration of emerging materials and device technologies capable of intrinsically generating spiking dynamics, including two-dimensional materials~\cite{lee2021artificial}, phase-change materials~\cite{xu2020recent}, and complex oxides~\cite{park2023complex}.

Among correlated oxides exhibiting an insulator-metal transition (IMT), vanadium dioxide (VO$_2$) has emerged as a promising candidate for analog neuromorphic networks due to its high-frequency resistive switching near room temperature (68 °C), device scalability, and endurance~\cite{yang2011oxide, zhao2025mott}. Its ability to sustain room-temperature self-oscillations under current bias with an additional capacitive load enables simple implementations of voltage-controlled oscillators~\cite{shukla2014synchronized,corti2020scaled,maher2024highly,delacour2023energy} and calibratable sensory neurons~\cite{yuan2022calibratable,fang2022artificial, qaderi2023millimeter}. Beyond individual devices, such oscillators can be coupled to form oscillatory neural networks (ONNs), which exploit the collective dynamics of nonlinear components. In these systems, information is encoded in the phase or frequency of oscillation, while computation emerges from synchronization phenomena within the network~\cite{hoppensteadt2000pattern,csaba2020coupled,delacour2021oscillatory,todri2024computing}. Compared with SNNs, ONNs rely on continuous-time phase dynamics, enabling efficient implementation of synchronization-based operations for optimization tasks, including classification~\cite{shukla2014pairwise,shukla2016ultra,dutta2019spoken}, Boolean satisfiability (3-SAT), and Ising-type problems~\cite{mostafa2015event,csaba2020coupled,dutta2021ising,maher2024cmos,maher2024highly}.

Although prior work has demonstrated CMOS-compatible VO$_2$ fabrication on SiO$_2$ substrates~\cite{vitale2015fabrication, corti2020scaled, maher2024highly, varini2025pulsed} and enhanced functionality in scaled hybrid VO$_2$-field-effect transistor (FET) architectures~\cite{shukla2015steep, shukla2014pairwise, qaderi2023millimeter, liu2024vo2}, functional 3D monolithic integration with FETs and scaling toward miniaturized one transistor-one memristor (1T-1MR) networks remain unexplored. In this work, we experimentally demonstrate the first 3D monolithic integration of VO$_2$ two-terminal devices with p-type silicon-on-insulator junctionless (SOI JL) FETs using a CMOS-compatible pulsed-laser deposition (PLD) process. By leveraging depletion-mode FETs for current-biased analog operation~\cite{doria2011junctionless}, this platform provides a scalable pathway toward very-large-scale oscillatory neural networks (ONNs) and back-end-of-line (BEOL) integrated spiking sensor arrays.

\section{Results and discussion}

Spiking neurons or oscillatory nodes are typically implemented in a one-transistor-one-memristor (1T-1MR) or one-resistor-one-memristor (1R-1MR) discrete architecture, where self-sustained oscillations are enabled by a DC voltage or current bias. We developed a scalable nanofabrication process for the systematic 3D integration of 1T-1MR oscillatory nodes, achieving an unprecedentedly small footprint and low parasitic losses. Our approach combines depletion-mode SOI JLFETs technology and VO$_2$ thin films. This choice is driven by the inherent suitability of junctionless technology for current-biased analog applications~\cite{doria2011junctionless}, and the low current required to induce an IMT compared to thick VO$_2$ films.  In the following, a description of the 3D integration scheme is presented.

\begin{figure}
    \centering
    \includegraphics[width=0.9\linewidth]{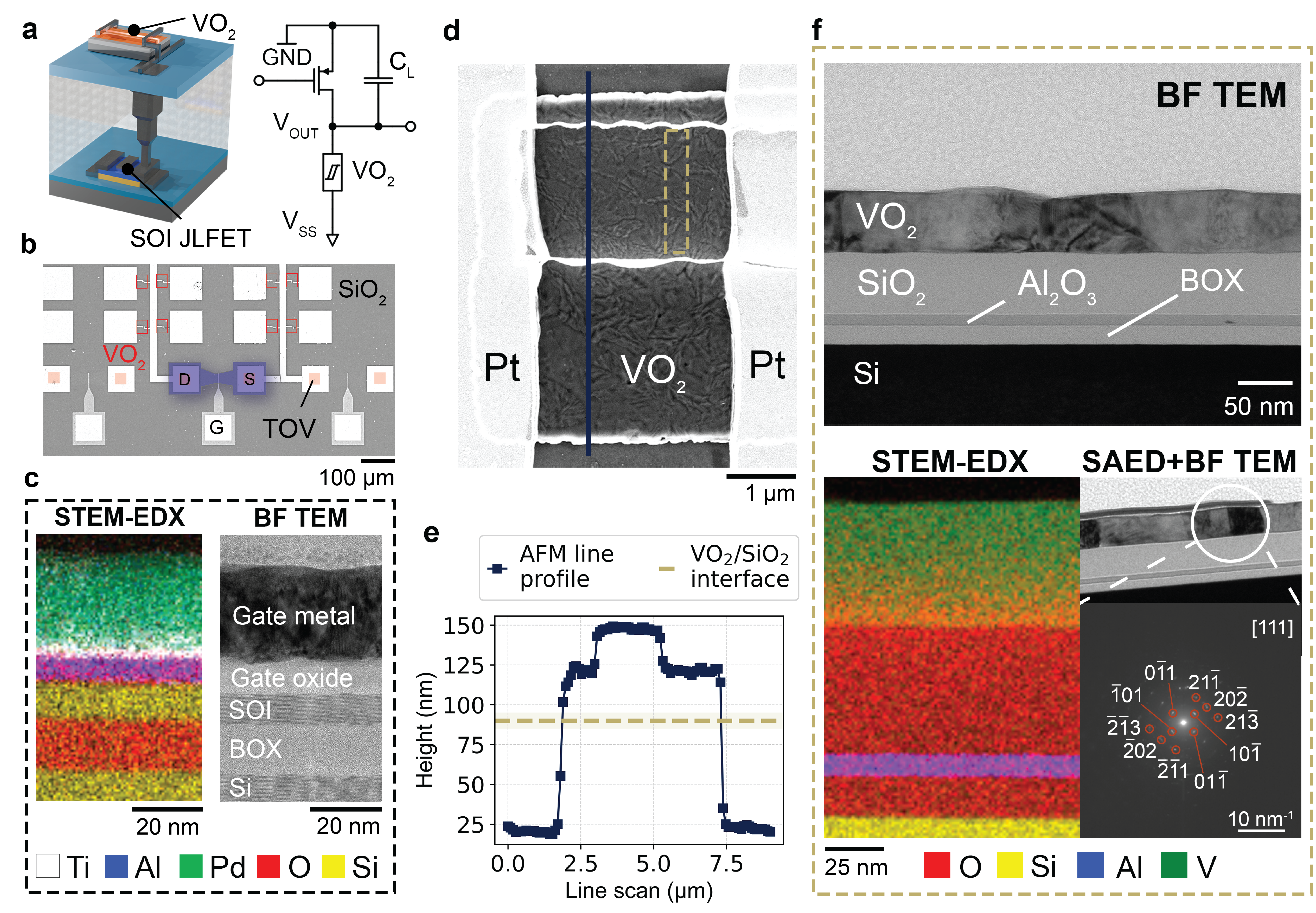}
    \caption{\textbf{Schematic of the spiking neuron comprising a VO$_2$ memristor and a JLFET, with corresponding material characterization.} \textbf{a} 3D-model of the 1T-1MR architecture and equivalent circuit schematic comprising an external capacitive load. VO$_2$ two-terminal devices are connected to the JL transistor through metal-filled through-oxide-vias (TOV). \textbf{b} SEM image of the fabricated chip, showing multiple JLFETs connected to the VO$_2$ node to enable IMT operation at the target current level. D, S, and G denote the drain, source, and gate terminals, respectively. \textbf{c} STEM-EDX and BF TEM images of the gate stack of the JLFETs after VO$_2$ integration. \textbf{d} SEM top view of a VO$_2$ nanosheet device. The blue line represents the direction of the line profile in (e) while the dashed-rectangle refers to the area of the lamella in (f). \textbf{e} Example of line profile extracted from an AFM topography measurement. \textbf{f} BF TEM image and STEM-EDX acquired in the active area of a VO$_2$ nanosheet, together with the SAED pattern and the corresponding BF TEM image of its acquisition region. The SAED pattern is indexed along the [111] zone axis of the VO$_2$ M1 phase with the corresponding reflections indicated.}
    \label{fig:Figure_1}
\end{figure}

\subsection{3D monolithic integration of VO$_2$ memristors and junctionless FETs}

A schematic of the integrated devices is reported in Fig.~\ref{fig:Figure_1}a: VO$_2$ two-terminal devices are connected to the drain of the transistors through a metal-filled via in a protective oxide layer. The chip layout consists of periodic arrays of VO$_2$ two-terminal devices connected to four p-type SOI JLFETs in parallel to precisely match the VO$_2$ threshold current at different temperatures. A scanning electron microscope (SEM) image of the fabricated devices is shown in Fig.~\ref{fig:Figure_1}b.

The SOI JL-FETs are fabricated on ultra-thin SOI channels of 15~nm, featuring a 20~nm-thick buried oxide (BOX). The estimated boron dopant concentration is $5\cdot10^{18}\;\mathrm{cm^{-3}}$, enabling full depletion of mobile carriers in the SOI channel for gate voltages above $\approx3$ V (with grounded substrate). The transistor gate stack includes 10~nm of Al$_2$O$_3$ as the gate dielectric and Ti/Pd (3/37 nm) as the gate metal. The effective metal workfunction and total interface traps per unit area were extracted from high- and low-frequency capacitance measurements, with values of 5.4~eV and $3\cdot10^{11}$~cm$^{-2}$, respectively~\cite{bersano2025integration}.

The detailed processes for SOI channel patterning, source and drain ohmic contacts, and JLFETs gate stack fabrication are included in Methods. The transistors were covered by 60~nm of SiO$_2$ deposited through atomic-layer deposition (ALD) at 300~°C to isolate the SOI active area from subsequent fabrication steps. A bright-field TEM (BF TEM) image of the channel cross-section and the corresponding energy-dispersive X-ray scanning transmission electron microscopy (STEM-EDX) map are shown in Fig.~\ref{fig:Figure_1}c, indicating no significative degradation of the SOI interfaces after VO$_2$ integration.

The VO$_2$ deposition was performed with pulsed laser deposition (PLD) from a V$_2$O$_5$ target in an O$_2$ atmosphere at a pressure of 6.6~mTorr, within a thermal budget not exceeding 430~°C (Supporting Information). The film was subsequently patterned by laser lithography and etched by ion beam etching (IBE) into nanosheet devices defining an active area of $6~\text{\textmu m}^2$. Ti/Pt contacts were then patterned on top by laser lithography, metal evaporation, and lift-off. A scanning electron microscopy (SEM) picture of a VO$_2$ device is reported in Fig.~\ref{fig:Figure_1}d. The active area of the nanosheet is surrounded by a thinner VO$_2$ residual layer to enforce resilience against structural modifications induced by high electric fields~\cite{conti2026electroforming}. Fig.~\ref{fig:Figure_1}e reports the line profile of a device channel acquired from an atomic force microscopy (AFM) topography map. The active area is 60~nm thick, while the surrounding residual layer has an average thickness of 30~nm. Notably, the thickness of the pristine PLD-VO$_2$ film was 100~nm and was subsequently reduced to 60~nm by IBE. This process produced a dense, thin film with large columnar grains, minimizing the risk of void formation in the channel~\cite{varini2025pulsed} while enabling device operation at currents below 30~$\text{\textmu A}$. A zoomed BF TEM image of the nanosheet active area, a STEM-EDX map and a selective area electron diffraction (SAED) pattern with its acquisition area are reported in Fig.~\ref{fig:Figure_1}f. The BF TEM analysis shows a structurally homogeneous VO$_2$ layer, with no significant rough regions observed on the top surface following thinning of the film. The VO$_2$ grains are columnar, with lateral dimensions ranging from 50~nm to 100~nm.
The STEM-EDX map highlights the presence of an oxygen gradient extending from the interface with SiO$_2$. Nonetheless, analysis of the SAED patterns indicates that the structure is consistent with the VO$_2$ M1 phase, owning to the precise stoichiometry control and parameters flexibility provided by the PLD technique. An example of a pattern indexed along the [111] zone axis is included in the figure. The detailed fabrication process of the VO$_2$ devices, including the ordered IBE steps is reported in Methods.

\subsection{Electrical characterization of junctionless FETs and VO$_2$ memristors}
\begin{figure}
    \centering
    \includegraphics[width=\linewidth]{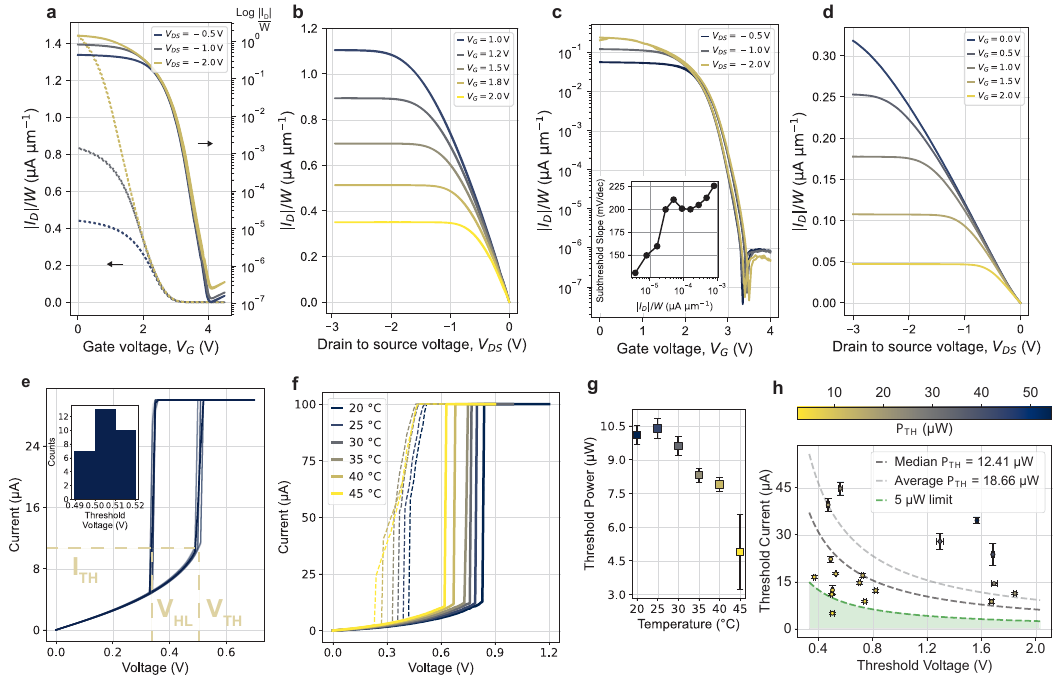}
    \caption{\textbf{Electrical characterization of JLFETs and VO$_2$ nanosheet memristors.} \textbf{a, b} Transfer and output characteristics of a JLFET prior to VO$_2$ integration. \textbf{c, d} Transfer and output characteristics of a JLFET after VO$_2$ integration. The inset in (c) shows the subthreshold slope as a function of the normalized drain current. The channel width and length are $W=30$~\textmu m and $L=2$~\textmu m, the Si substrate is at 0~V. \textbf{e} Example of I-V characteristic for a VO$_2$ nanosheet device together with the indication of the upper threshold voltage ($V_{TH}$), the holding voltage ($V_{HL}$), and current ($I_{TH}$). The darker curve is the average of 30 consecutive measurements (in light blue) obtained with a voltage step of 10~mV. The histogram of $V_{TH}$ is reported in the insert. \textbf{f} I-V characteristics measured at different temperatures. \textbf{g} Trend of the threshold power $P_{TH}$ with the temperature for the device shown in (f). Each point is an average of 30 consecutive measurements. \textbf{h} Room temperature distribution of threshold voltage and current for the measured devices, with the corresponding power represented by the top color bar. Dashed lines indicate the median and mean power computed across all devices. Over 50\% of the observed thresholds are concentrated within a narrow low-power cluster.}
    \label{fig:Figure_2}
\end{figure}

The electrical characterization of an SOI JL-FET and the quasi-static voltage sweep measurements of the fabricated VO$_2$ devices are shown in Fig.~\ref{fig:Figure_2} for some representative devices. Figs.~\ref{fig:Figure_2}a-d show the room temperature transistor transcharacteristics and bias curves before and after completing the VO$_2$ integration. As evidenced by the negligible dependence of the drain current ($I_D$) on the bias voltage ($V_{DS}$) in the subthreshold regime and the linearity of the $I_D$-$V_{DS}$ characteristics at low drain bias, the source and drain contacts remain ohmic before and after PLD annealing, despite a tenfold increase in contact resistance. The total series resistance after PLD annealing, extracted from the output characteristics, is $R_{\mathrm{tot}}\approx343$~k$\Omega$. The source/drain contact resistance was then determined from the intercept of the $R_{\mathrm{tot}}$ versus channel length plot, yielding $R_{SD}\approx331~$k$\Omega$ for devices with channel lengths ranging from 2 to 5~\textmu m (Supporting Information).

The effective holes mobility was extracted in accumulation regime using a modified Y-method ($Y=\frac{I_D}{gm^{1/2}}$, where $g_m$ is the transconductance) to account for the effect of bulk conduction~\cite{jeon2013revisited}. For the long-channel devices (W~$=5$~\textmu m), the extracted mobility is $\mu_{\mathrm{eff}}=49.7$~cm$^2$/Vs before annealing and $\mu_{\mathrm{eff}}=49.4$~cm$^2$/Vs after annealing in the PLD system, indicating a negligible impact of the thermal process. The effective mobility exhibits only a weak dependence on substrate temperature up to 50~$^\circ$C (see Supporting Information). This behavior is consistent with the characteristics of junctionless transistors, which typically exhibit lower transconductance than inversion-mode devices, leading to reduced flicker noise ($\propto g_m^2$) and a weaker temperature dependence of $g_m/I_D$, associated with reduced mobility degradation~\cite{doria2011junctionless}.

Fig.~\ref{fig:Figure_2}e shows a typical I-V characteristic of VO$_2$ two-terminal devices as average of 30 consecutive measurements obtained with a 10 mV voltage step, with the insert capturing the distribution of the upper threshold voltage ($V_{TH}$) for the different runs. We refer to the holding voltage ($V_{HL}$) as the lower threshold for the transition from the metallic to the insulating state. The nanosheets devices exhibit $<$ 500 mV hysteresis windows, and a low-stochasticity, with the highest observed $V_{TH}$ coefficient of variation being about $\sigma/\mu = 3\%$ (with $\sigma$ and \textmu~the standard deviation and mean value), owing to an optimized combination of film structure and device design. An example of the temperature dependence of the VO$_2$ electrical behavior is reported in Fig.~\ref{fig:Figure_2}f. Increasing the external temperature results into a noticeable decrease of $V_{TH}$ and $V_{HL}$, with a more moderate variation on $I_{TH}$, leading to an overall reduction of the threshold power $P_{TH}$ (Fig.~\ref{fig:Figure_2}g). Notably, at room temperature, more than 50\% of the measured devices fall within a narrow low-power cluster (Fig.~\ref{fig:Figure_2}h) characterized by sub-1 V $V_{TH}$ and an average $P_{TH}$ of about 8~\textmu W, with the lowest measured value approaching 2.5~\textmu W. The observed variations in the threshold parameters are mainly attributed to the intrinsic resistivity variability of polycrystalline films, as well as to process-related differences and metal lift-off yield that affect the effective channel length. 

A generic device was modeled in COMSOL Multiphysics to investigate power scaling relative to the tested geometry with W$_0$ = 2 \textmu m and L$_0$ = 3 \textmu m (see Methods). The finite-element modeling (FEM) simulation returns a $P_{TH}$ close to 17~\textmu W, in agreement with the average of the experimentally measured data. Notably, the model predicts sub-5~\textmu W power consumption through moderate scaling of W or L (Supporting Information). In these simulations, only the device active area was varied, while its thickness was not scaled. Indeed, excessively thin film are more prone to electroforming, thus device failure in-operation, compared to thicker layers with comparable on/off resistance ratio ($\frac{R_{20^\circ C}}{R_{95^\circ C}}$)~\cite{conti2026electroforming}. In our nanosheet devices, the on/off resistance ratio is on the order of tens, which is sufficient to ensure clear separation between the two conductance states and reduce the risk of electrical breakdown. 

\subsection{Deterministic and stochastic voltage spiking operation}
\subsubsection{Current and temperature dependence of spiking dynamics in integrated neuristors}

\begin{figure}
    \centering
    \includegraphics[width=0.5\textwidth]{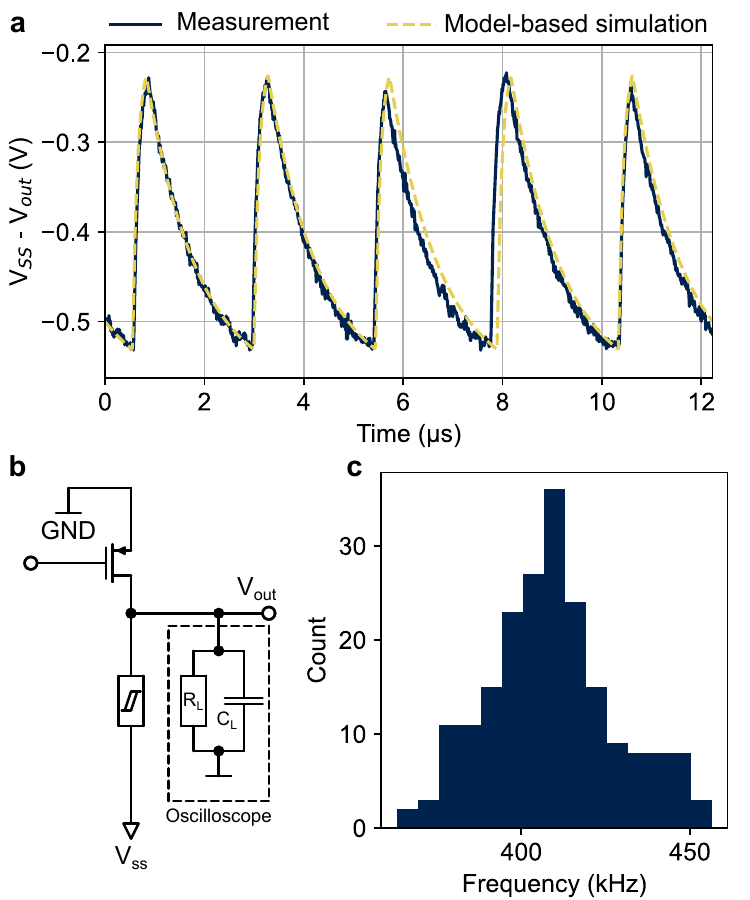}
    \caption{\textbf{Self oscillations and frequency histogram of a fully integrated 1T-1MR artificial neuron.}
    \textbf{a} Measured waveform at room temperature, showing a minimum VO$_2$ energy consumption of $\approx$18~pJ per spike (Methods). The dashed line represents the oscillation reproduced by the model in Eq.~\ref{Eq:frequency_model} with the following parameters: $V_{TH}$=531~mV, $V_{HL}=$225~mV, $R_i=$21.1~k$\Omega$, $R_m=$1.98~k$\Omega$, $V_{oi}=$166.5~mV, $V_{om}=$168~mV, $C_L = 50$~pF. \textbf{b} Circuit schematic of the device, including the capacitive ($C_L \approx 50$~pF) and resistive ($R_L \approx 1$~M$\Omega$) loads given by the oscilloscope. \textbf{c} Histogram of the oscillation frequency corresponding to the measurement in (a), obtained over more than 200 cycles. The peak frequency is 410~kHz, with a standard deviation of $\sigma = 18.8$~kHz.}
    \label{fig:Figure_3}
\end{figure}

The integrated neuristors were tested with an estimated capacitive load of $C_L \approx 50$~pF, corresponding to the combined input capacitance of the coaxial cable and the oscilloscope used for electrical characterization. Self-sustained oscillations up to 410~kHz were measured at room temperature (Fig.~\ref{fig:Figure_3}), corresponding to a minimum energy consumption of 18~pJ per spike (obtained by voltage integration) and an average memristor power dissipation of 8~\textmu W. In agreement with the narrow hysteresis window shown in Fig.~\ref{fig:Figure_2}e, the oscillations exhibit a peak-to-peak voltage of approximately 300~mV, on the same order of magnitude as biological action potentials~\cite{hodgkin1952quantitative}. Notably, compared to previous works based on epitaxial VO$_2$ \cite{shukla2014synchronized, dutta2021ising} and higher annealing temperatures~\cite{yuan2022calibratable, maher2024highly, corti2020scaled}, our nanosheet PLD thin film devices enable spiking operation with x10 decrease in energy per cycle.

\begin{figure}
    \centering
    \includegraphics[width=0.7\linewidth]{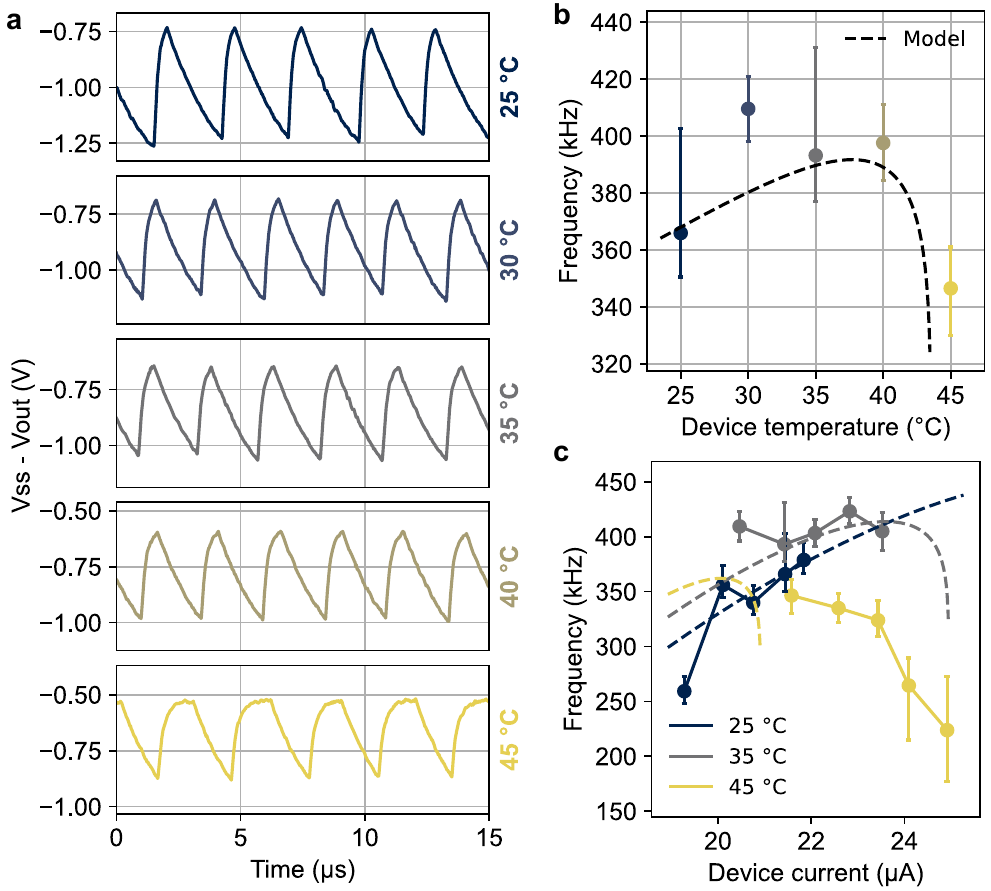}
    \caption{\textbf{Dependence of integrated VO$_2$ oscillators on device temperature and bias current.} \textbf{a} Measured waveforms at different substrate temperatures under a constant bias current of 21.46~\textmu A. \textbf{b} Extracted oscillation frequency at different substrate temperatures. The bias current is fixed at 21.46~\textmu A; Error bars represent the interquartile range, spanning from the first quartile (Q1) to the third quartile (Q3). \textbf{c} Dependence of the oscillation frequency on the bias current at three selected temperatures. The non-monotonic behavior in (b) and (c) is explained by the analytical model (Eq.~\ref{Eq:frequency_model}), shown in dashed line. The deterministic model no longer reproduces the experimental data at temperatures above 45~$^\circ$C, where a stochastic analysis is required. }
    \label{fig:Figure_4}
\end{figure}

We then investigated the tunability of the oscillator frequency by characterizing the analog spiking neurons as a function of substrate temperature and transistor bias current, thereby evaluating drift effects from chip self-heating and ambient temperature variations. Representative waveforms acquired under constant-current bias at temperatures ranging from 25~$^\circ$C to 45~$^\circ$C are shown in Fig.~\ref{fig:Figure_4}a. The measurements reveal a non-monotonic dependence of the oscillation frequency, which can be captured by a compact analytical model. Unlike previously reported electrothermal formulations~\cite{zhang2025device}, our model approximates the nonlinear threshold-switching characteristic using piecewise-linear branches and derives the oscillation period by analytically solving the load-capacitor charging dynamics between the two switching thresholds. 

The oscillation period $p$ is expressed as the sum of a \textit{rising} and a \textit{falling} contribution (denoted by the subscripts $r$ and $f$), both governed by the charging dynamics of the load capacitor $C_L$. Denoting by $V_a$ the asymptotic voltage across the VO$_2$ device reached after a switching event under constant bias, and by $I$ the bias current, the oscillation period can be written as:

\begin{equation}
    p = \tau_r \ln\left(\frac{V_{HL}-V_{ar}}{V_{TH}-V_{ar}}\right) + \tau_f \ln\left(\frac{V_{TH}-V_{af}}{V_{HL}-V_{af}}\right)
    \quad \text{with} \quad
    \begin{cases}
        V_{ar}=V_{oi}+R_{i,eq}I, \quad \tau_r=C_LR_i\\
        V_{af}=V_{om}+R_{m,eq}I, \quad \tau_f=C_LR_m
    \end{cases}
    \label{Eq:frequency_model}
\end{equation}

where the model parameters and their extraction procedure are summarized in Table~\ref{tab:model_parameters}. The parameters were extracted by fitting the exponential rising and falling phases of the voltage across the VO$_2$ device over a range of bias currents and temperatures (see Methods for details).

\begin{table}[h]
\centering
\caption{Parameters of the analytical model and extraction procedure.}
\label{tab:model_parameters}

\begin{tabularx}{\textwidth}{%
>{\raggedright\arraybackslash}p{0.16\textwidth}
>{\raggedright\arraybackslash}p{0.27\textwidth}
>{\raggedright\arraybackslash}X}
\hline
\textbf{Parameter} & \textbf{Physical meaning} & \textbf{Extraction procedure} \\
\hline

$C_{L}$ &
Estimated capacitive load &
Calculated from the combined impedance of the coaxial cable and the oscilloscope input channel ($\approx$ 50 pF). \\

$R_{i,eq}$ &
Equivalent resistance in the insulating state &
Obtained from the fitted rising-phase time constant $\tau_r$. \\

$R_{m,eq}$ &
Equivalent resistance in the metallic state &
Obtained from the fitted falling-phase time constant $\tau_f$. \\

$V_{oi}$ &
Offset voltage in the insulating state &
Calculated from the fitted asymptotic voltage of the rising phase ($V_{oi}=V_{ar}-IR_{i,eq}$). \\

$V_{om}$ &
Offset voltage in the metallic state &
Calculated from the fitted asymptotic voltage of the falling phase ($V_{om}=V_{af}-IR_{m,eq}$). \\

$V_{TH}$ &
Threshold voltage &
Extracted as the maximum voltage of each oscillation cycle. \\

$V_{HL}$ &
Holding voltage &
Extracted as the minimum voltage of each oscillation cycle. \\

\hline
\end{tabularx}

\vspace{0.4em}

\begin{minipage}{\textwidth}
\footnotesize
\textit{Note:} From a circuit perspective, the VO$_2$ memristor is modeled as a linear resistor ($R{i,eq}$ or $R_{m,eq}$) in series with a voltage source ($V_{oi}$ or $V_{om}$) to approximate the nonlinear I-V characteristic, as schematically depicted in the Supporting Information.
\end{minipage}

\end{table}

The current and temperature dependence of each parameter is provided in the Supporting Information. $V_{TH}$ and $V_{HL}$ exhibit negligible dependence on bias current but decrease with increasing temperature. The same trend is observed for the offset voltages $V_{oi}$ and $V_{om}$, whereas the fitted values of $R_{i,eq}$ and $R_{m,eq}$ show a weak dependence on bias current, varying by less than 15.8\% and 13.5\%, respectively, over the entire measurement range. Remarkably, the temperature dependence of all parameters can be described by first-order polynomials over the 25-45~$^\circ$C temperature range.

For a given temperature and bias point, Eq.~\ref{Eq:frequency_model} directly identifies the two conditions required for sustained oscillation in an ideal system: $V_{ar}>V_{TH}$ and $V_{af}<V_{HL}$. The model explains the intrinsically non-monotonic dependence of the oscillation frequency on device current: as the asymptotic rising and falling voltages approach their respective threshold and holding voltages, the corresponding charging times increase, resulting in a lower oscillation frequency. Because the VO$_2$ hysteresis window is itself temperature dependent, the model also predicts a non-monotonic dependence of oscillation frequency on temperature (Fig.~\ref{fig:Figure_4}b). This simple yet insightful analytical model provides a practical framework for designing spiking analog circuits by predicting the maximum oscillation frequency achievable for a given hysteretic window and thin-film parameters, while guiding the selection of biasing conditions that account for self-heating effects during CMOS operation. While the experimental results are reproduced for temperatures below 45~$^\circ$C (Fig.~\ref{fig:Figure_4}c), the dataset acquired at the highest temperature falls outside the oscillation regime defined by Eq.~\ref{Eq:frequency_model}, and the switching dynamics become dominated by stochastic triggering, as analyzed in the following.

\subsubsection{Stochastic dynamics of integrated VO$_2$ spiking neuristors}

\begin{figure}
    \centering
    \includegraphics[width=\linewidth]{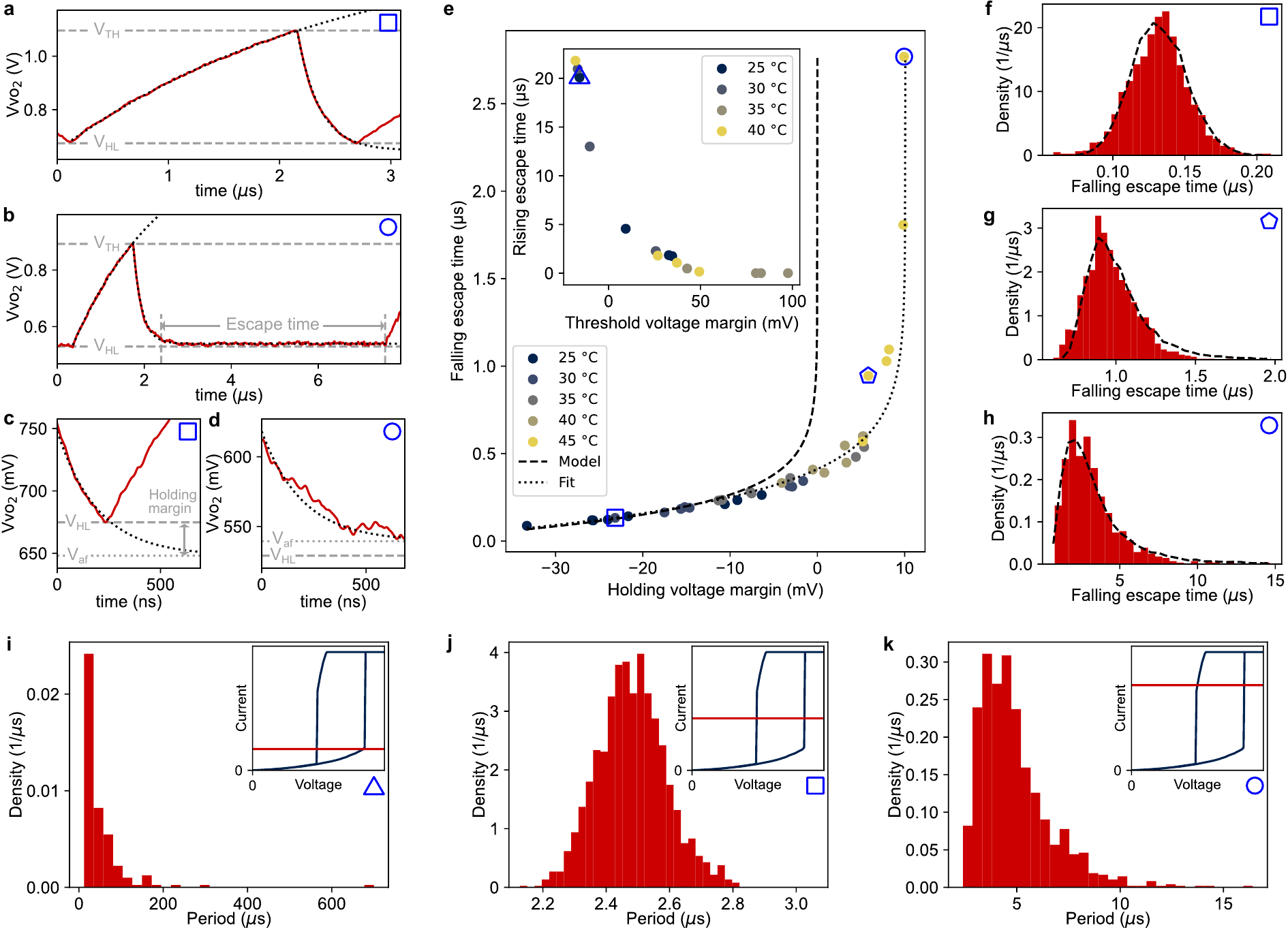}
    \caption{\textbf{Analysis of firing stochasticity in integrated VO$_2$ spiking neurons.} 
\textbf{a} Representative spike cycle for a circuit operating within the oscillation regime. 
\textbf{b} Representative spike cycle for a circuit operating outside the analytically defined holding-voltage oscillation condition. 
\textbf{c} Close-up view of (a). \textbf{d} Close-up view of (b).
\textbf{e} Falling escape time as a function of the holding voltage margin (each point represents the median over at least 1000 cycles). Inset: rising escape time as a function of the threshold voltage margin for a different device (median over at least 200 cycles).
\textbf{f–h} Falling escape time histograms for different holding voltage margins, corresponding to the points highlighted in (e).
\textbf{i–k} Period histograms for different oscillation regimes, corresponding to the points highlighted in (e). Insets provide a graphical representation for illustrative purposes only. }
\label{fig:Figure_5}
\end{figure}

Representative time traces under bias conditions inside and outside the deterministic oscillation regime are shown in Fig.~\ref{fig:Figure_5}a,b, with zoomed-in views in Fig.~\ref{fig:Figure_5}c,d. When biasing the circuit within the oscillation condition, the asymptotic falling voltage ($V_{af}$) remains below the holding voltage $V_{HL}$, ensuring the triggering of the next cycle (Fig.~\ref{fig:Figure_5}c). Conversely, outside this regime, $V_{af}$ exceeds $V_{HL}$, preventing immediate retriggering (Fig.~\ref{fig:Figure_5}d). As a result, a resting (escape) time emerges between successive spiking events, as clearly observed in Fig.~\ref{fig:Figure_5}b. In this regime, the next spike is no longer deterministically generated but is instead stochastically triggered by voltage noise.

The difference between $V_{af}$ and $V_{HL}$, which defines the oscillation condition, is referred to as the holding voltage margin, and an analogous definition applies to the threshold voltage margin. We define the rising and falling escape times as the time between the signal reaching 50~mV from the threshold or holding voltage and the start of the next cycle. This parameter definition implicitly accounts for the effects of voltage noise in the analytical description of the charging and discharging times. Fig.~\ref{fig:Figure_5}e shows the dependence of the falling escape time on the holding voltage margin (median values over all cycles are reported), where the dashed line corresponds to the analytical model prediction. A fit to the experimental data reproduces the trend predicted by the analytical model, with an offset of approximately 10~mV, consistent with the noise level of the measurement setup. The dependence of the rising escape time on the threshold voltage margin for a different device is shown in the inset of Fig.~\ref{fig:Figure_5}e, and consistent dependencies are observed across several devices for both oscillation thresholds, supporting an intrinsic physical origin of this behavior (Supporting Information).

Figs.~\ref{fig:Figure_5}f-h present histograms of the falling escape time, along with corresponding statistical simulations (Methods and Supporting Information), for different holding voltage margins highlighted in Fig.~\ref{fig:Figure_5}e. In Fig.~\ref{fig:Figure_5}f, the falling escape time distribution exhibits a near-Gaussian profile, as expected within the oscillation regime. This behavior can be mostly attributed to stochastic fluctuations in the holding voltage, as observed in the quasi-static I-V curves. These fluctuations were independently characterized and incorporated into the statistical simulations. An additional $1/f$ voltage noise was then included to reproduce the experimental distribution. As the holding-voltage margin increases, the signal slope decreases, leading to a progressively skewed escape-time distribution, and the histogram shifts toward shorter escape times (Fig.~\ref{fig:Figure_5}g). We attribute this behavior to a time-dependent variation of the holding voltage caused by thermal dissipation in the substrate, which is modeled in the simulations as an equivalent thermal RC delay. At even larger holding-voltage margins, the escape-time distribution approaches a Poisson-like distribution (Fig.~\ref{fig:Figure_5}h), consistent with a noise-driven triggering mechanism.

Finally, Figs.~\ref{fig:Figure_5}i-k show the distribution of the oscillation period under different current biases: (i) outside the threshold-defined oscillation condition, (j) within the oscillation regime, and (k) outside the holding-defined oscillation condition. The period distributions follow exponential, Gaussian, and Gamma distributions, as also observed in sub-species of biological neurons~\cite{liu2021tantalum}. These measurements on fully integrated VO$_2$ devices confirm the experimental results and interpretation of the firing dynamics previously discussed in TaS$_2$-based~\cite{liu2021tantalum} and FET-based integrated neuristors~\cite{maeng2025reconfigurable}. In particular, for our nanosheet VO$_2$ devices, a low $R_i/R_m$ ratio enables the statistical analysis of the oscillation behavior near the holding boundary while maintaining moderate power consumption. This analysis provides new insights into the design of fully integrated thin-film VO$_2$ devices for stochastic applications, such as probabilistic computing and phase-transition-based random number generators~\cite{jerry2017stochastic}.

\subsubsection{Voltage-controlled oscillations and spike-rate encoding}

\begin{figure}
    \centering
    \includegraphics[width=\linewidth]{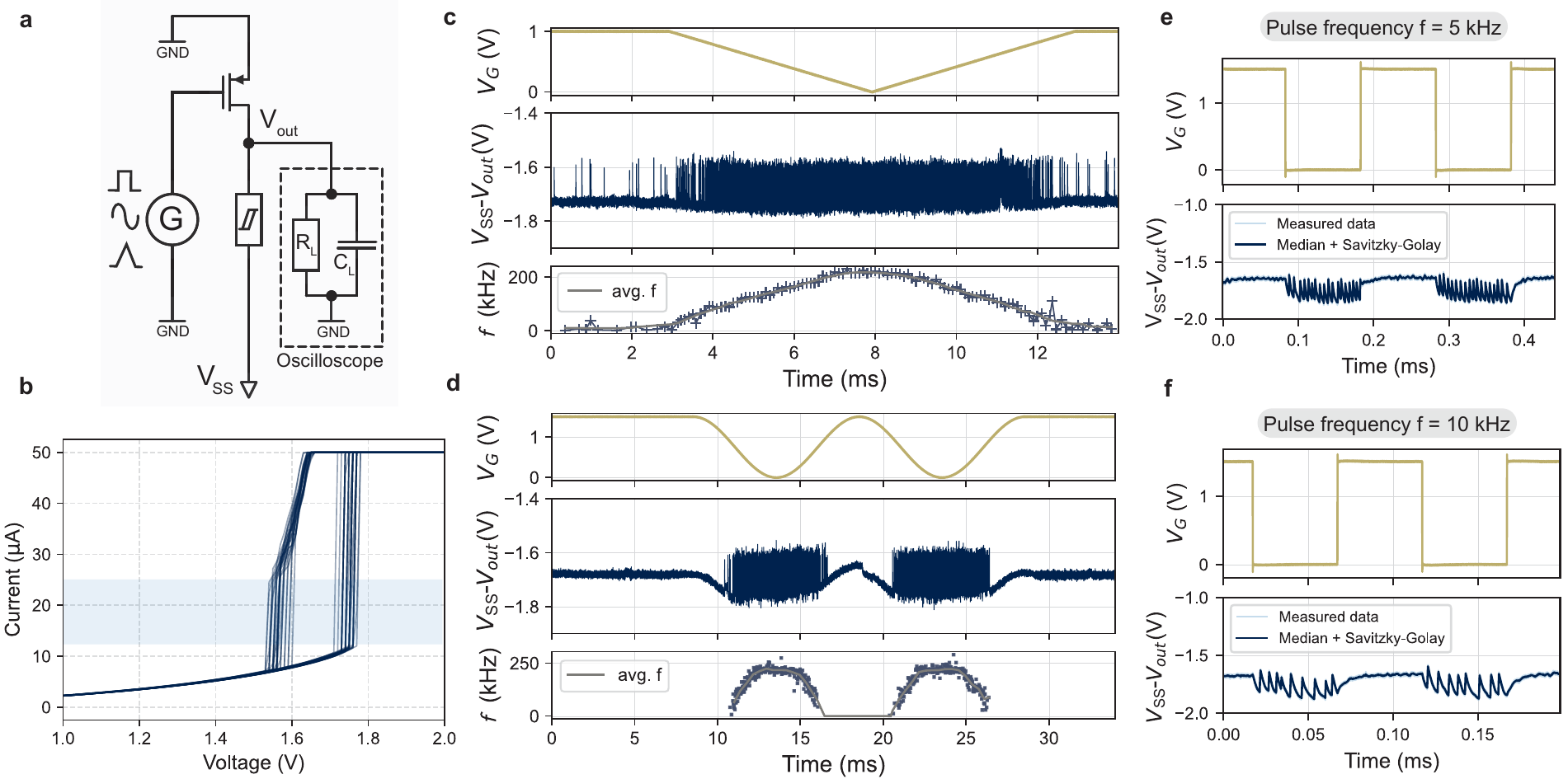}
    \caption{\textbf{Dynamic operation of an integrated VO$_2$ voltage-controlled oscillator.} \textbf{a} Circuit schematic of the integrated device, including the external voltage function generator and the equivalent load representing the oscilloscope channel. The load resistance and capacitance are R$_L$=1~M$\Omega$ and C$_L$$\approx$50~pF, respectively. \textbf{b} Static I-V characteristic of the device under test, showing 30 cycles measured at 25~$^\circ$C. The light-blue area indicates the bias-current range swept by the transistor under gate-voltage control. \textbf{c,d} Measurements obtained by applying a linear ramp and a sinusoidal voltage signal to the gate, respectively, showing both the raw oscilloscope traces and the extracted evolution of the output oscillation frequency. \textbf{e,f} Oscillator response to square gate pulses at 5~kHz and 10~kHz, respectively. The number of measured output spikes increases linearly with the gate-pulse duration, demonstrating the operating principle of spike-rate encoding.}
    \label{fig:Figure_6}
\end{figure}

\begin{figure}
    \centering
    \includegraphics[width=\linewidth]{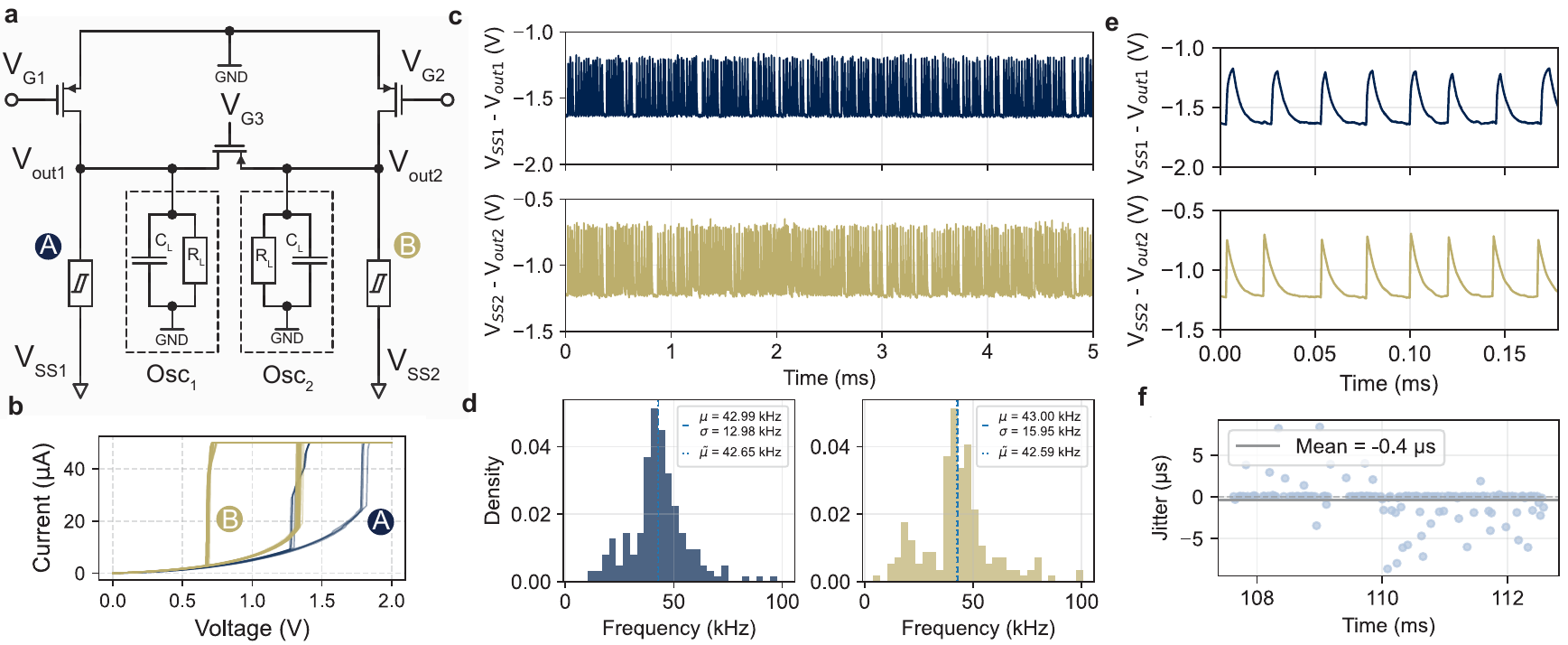}
    \caption{\textbf{On-chip active resistive coupling of integrated VO$_2$ oscillators.} \textbf{a} Circuit schematic of two 3D-integrated VO$_2$ oscillators (A and B) coupled via an on-chip depletion-mode JLFET. The schematic includes the equivalent loads corresponding to the two oscilloscope channels, with R$_L$ = 1~M$\Omega$ and C$_L \approx$ 50~pF. \textbf{b} Static I-V characteristics of the devices under test, measured at 25~$^\circ$C. \textbf{c} Measured waveforms over a time scale of 5~ms, along with the extracted frequency (raw data and moving average) for both channels. The average frequency for the two channels is 43~kHz, with a standard deviation of $\sim$13~kHz and $\sim$15~kHz. \textbf{d} Zoom-in of the measured waveforms over a time scale of 200~\textmu s, showing in-phase oscillations with the same frequency. \textbf{e} Computed time evolution of the jitter (\textit{i.e.}, the delay between voltage spikes of the two channels) over the measurement shown in (c).}
    \label{fig:Figure_7}
\end{figure}

The operation of an integrated VO$_2$ device as a dynamic voltage-controlled oscillator (VCO) is shown in Fig.~\ref{fig:Figure_6}. By applying time-varying voltage signals to the transistor gate, we emulate an analog CASN configuration (or sensory neuron)~\cite{tan2020tactile,yuan2022calibratable}, in which voltage fluctuations typically associated with a potentiometric transduction mechanism are translated into variations in the output oscillation frequency. Three different signals were applied to the gate of a p-JLFET using an external function generator: a linear ramp, a sinusoid, and digital square pulses with varying frequency and a 50$\%$ duty cycle, as schematically depicted in Fig.~\ref{fig:Figure_6}a. The triangular voltage sweep in Fig.~\ref{fig:Figure_6}c highlights an almost linear dependence of the oscillation frequency on the gate voltage ($\mathrm{V_G}$) when the transistor is biased at the boundary between the linear and saturation regimes. This behavior results in periodic frequency modulation under a sinusoidal drive, as shown in Fig.~\ref{fig:Figure_6}d, and underlies the operation of firing-rate coding in an analog neuron. 

For digital pulse inputs (Figs.~\ref{fig:Figure_6}e-f), the output exhibits deterministic oscillatory bursts, where each oscillation cycle is interpreted as a discrete spike event. The number of oscillation cycles increases proportionally with pulse duration, resulting in a doubling of spike count when the input frequency changes from 5~kHz to 10~kHz. This demonstrates pulse-duration-to-spike-count conversion, consistent with spike-rate encoding mechanisms used in rate-based spiking neural networks~\cite{fida2023active, guo2021neural} and event-driven neuromorphic sensing systems~\cite{yi2018biological, yuan2023neuromorphic}.

\subsubsection{On-chip active resistive coupling of integrated VO$_2$ nano-oscillators}

Physical computing with ONNs typically exploits the rich collective dynamics of coupled oscillators, which span regimes ranging from chaotic behavior to synchronization. The latter is particularly relevant, as it reflects the stability and intrinsic memory of a network of coupled dynamical systems, where electrical coupling is commonly achieved using resistors or capacitors~\cite{todri2024computing}.

Following the characterization of the tunability of the oscillation frequency with current bias and its temperature dependence, we investigated the synchronization of two VCOs monolithically integrated on the same substrate. The coupling is mediated by the resistive channel of a p-JLFET connecting the oscillating nodes. When the junctionless transistors are biased in deep accumulation at low $V_{DS}$, their channel can be modeled by an equivalent series resistance of 343~k$\Omega$. This configuration therefore mimics previous studies on the synchronization dynamics of discrete VO$_2$ oscillators coupled through an external resistor~\cite{corti2020time}. The circuit schematic of the experimental setup and the static I-V characteristics of the two VO$_2$ devices are shown in Fig.~\ref{fig:Figure_7}a,b. The I-V curves reveal a significant mismatch between the two VO$_2$ devices, which were selected to represent the largest on-chip mismatch measured in this work. To compensate for the different voltage windows of the two devices, an asymmetric $V_{SS}$ bias was applied to the left and right branches of the circuit ($V_{SS1}$ = -11.5~V, $V_{SS2}$ = -10~V), resulting in different bias currents for the two VCOs. 

Self-sustained oscillations measured by the oscilloscope at the two output nodes ($V_{out1}$ and $V_{out2}$ in the circuit schematic) are shown in Fig.~\ref{fig:Figure_7}c over a time interval of 5~ms. For both oscillators, the oscillation amplitudes are consistent with the voltage windows shown in the static I-V characteristics. When the three junctionless transistors are tuned to deep accumulation ($V_{G1}$, $V_{G2}$, $V_{G3}<0$~V), the voltage oscillations at the left ($V_{out1}$) and right ($V_{out2}$) nodes become linearly coupled, and the two VO$_2$ devices converge to a common collective oscillation frequency with mean value $\mu = 43$~kHz. Fig.~\ref{fig:Figure_7}d shows the histograms of the extracted frequencies for the two channels over the full 5~ms time window. The mean and median frequencies of the two waveforms differ by 0.017\% and 0.14\%, respectively, while the different standard deviations reflect the distinct bias points at which the two devices traverse their respective I-V curves.

A zoom-in over a 200~\textmu s time interval is shown in Fig.~\ref{fig:Figure_7}e, it highlights the temporal synchronization of the voltage spikes in the two waveforms. The phase shift between the two measured signals remains nearly constant throughout the full time window shown in Fig.~\ref{fig:Figure_7}c, as illustrated by the jitter time-evolution in Fig.~\ref{fig:Figure_7}f. Here, jitter is defined as the time delay between the starting points of corresponding voltage spikes measured at the two output nodes.

This proof-of-concept result establishes the feasibility of extending integrated VO$_2$ oscillators from single-device operation toward cooperative oscillatory dynamics on-chip by means of tunable junctionless FETs. Future studies should investigate the tuning of different phase relationships as a function of coupling strength (R$_{Coupling}$ as a function of a control voltage) to achieve adaptive synchronization in networks of coupled oscillators, thereby emulating synaptic potentiation and depression in phase-based ONNs.

\section{Conclusions}

This work demonstrates the first integrated 1T-1MR VO$_2$-based oscillators achieved through stack engineering of an ultra-thin-body, highly doped SOI substrate.

Previous integrated architectures based on functional oxides have explored different material systems~\cite{coll2019towards}. Among these, a subset of studies relies on TaO$_x$-based memristors directly integrated on CMOS circuits to implement multi-state memory and oscillator functionalities, exploiting bistability arising from the temporary formation and dissolution of conductive filaments~\cite{sharma2015high, sheng2019low, golonzka2019non}. In contrast, the polycrystalline PLD VO$_2$ nanosheet devices engineered in our study exhibit a fully reversible, current-driven IMT phase transition. This switching mechanism is not solely governed by ionic motion (defects) and typically results in better reversibility of the conductance states than filamentary resistive switching approaches (for physical insights into the percolation dynamics in VO$_2$ films, see Tiwari et al.~\cite{tiwari2026near}). 

Quasi-static and dynamic measurements were performed from 25~°C up to 45~$^\circ$C, revealing voltage-controlled oscillatory dynamics and spike thresholding behavior that had previously been reported only in discrete-component VO$_2$ oscillators. The compact devices achieve room-temperature frequency tunability from 40~kHz to 410~kHz, with sub-10~\textmu W power consumption in the VO$_2$ device.

The dependence of the oscillation frequency on device temperature and bias current was experimentally characterized and accurately captured through analytical fitting and circuit simulations. The results show an almost linear dependence of both threshold and holding voltages on device temperature, as well as a non-monotonic frequency evolution, which is governed by the time constants associated with the insulating and metallic resistances of the thin film. These experimental studies provide new insights into the design of fully integrated oscillating systems operating under varying temperature and bias conditions.

The analysis of the stochastic spiking shows an evolution of the escape-time and period distributions from exponential to Gaussian and back to exponential when sweeping across the hysteresis window from the threshold to the holding voltage. This analysis was enabled by the relatively low resistive ratio of the fabricated nanosheet devices, which allows to explore the oscillation regime beyond the holding-defined range at bias current levels compatible with transistor operation, thereby opening new possibilities for low-power randomness-based devices such as fully-integrated random number generators.

Device operation as a voltage-controlled oscillator was demonstrated using linear ramp, sinusoidal, and square pulse waveforms. Moreover, for the first time, resistively coupled oscillatory dynamics were experimentally realized using an integrated JLFET as an active coupling element. 

Further investigations of device-to-device and die-to-die variability in the integrated oscillators will be required to optimize the process for very large-scale integration. Nevertheless, this proof-of-concept experimental study highlights the feasibility of 3D integration of VO$_2$ devices with CMOS platforms within a More-than-Moore (\textit{i.e.}, CMOS+X) framework, and opens a pathway toward fully integrated systems for oxide electronics, neuromorphic sensing, and computing.

\newpage
\section{Methods}
\subsection{Fabrication of junctionless FETs and VO$_2$ memristors}
\textbf{UTB-SOI JLFETs - }The SOI substrates were implanted with boron via ion-beam implantation, targeting a dopant concentration of $5\cdot10^{18}~\mathrm{cm^{-3}}$. The SOI was then patterned by maskless laser writing (Heidelberg MLA 150) using AZ 3007 photoresist, followed by deep reactive ion etching (Alcatel AMS 200 SE system). A 10~nm layer of Al$_2$O$_3$ (gate oxide) was deposited by ALD at 300~°C using a Beneq TFS200 system with TMA and H$_2$O precursors. Ti/Pd front gates were subsequently defined by maskless lithography, followed by e-beam evaporation (Leybold Optics LAB600H) and lift-off. PtSi contacts were fabricated by e-beam evaporation and lift-off of Pt, followed by rapid thermal annealing in forming gas at 350~°C for 15 minutes. Finally, the chip was encapsulated with 70~nm of ALD SiO$_2$ deposited at 300~°C. The electrical characterization shown in Fig.~\ref{fig:Figure_2}(a-b) was performed prior to this final encapsulation step.\\
\textbf{VO$_2$ two-terminal devices - }VO$_2$ was deposited with a PLD system (Solmates SMP 800) from a V$_2$O$_5$ target under an oxygen partial pressure of 6.6 mTorr, at a baseline temperature of 400 °C (reaching a maximum temperature of 430~°C). The film was thinned down from 100 nm to 60 nm with a IBE (Veeco Nexus IBE350) system. Devices (active area and residual layer) were patterned by direct laser writing (Heidelberg MLA 150) using a AZ 1512 HS photoresist, and subsequently etched with IBE. The active area of the nanosheets was patterned by electron-beam lithography (Raith EBPG 5000+) using HSQ-FOX16 resists, and it was etched with IBE, followed by BHF-dip for resist-stripping. A schematic of the sequence of the IBE etching steps is reported in the Supporting Information for clarity. Through-oxide-vias connecting the VO$_2$ layer to the JLFET drain were patterned by direct laser writing using AZ 1512 HS photoresist. The vias were opened by wet etching in buffered-HF and filled with e-beam evaporated Ti/Pt (5~nm/55~nm, Leybold Optics LAB600H tool). Contacts were patterned by direct laser writing with LOR5A+AZ 1512 HS double-resist layer for metals lift-off.

\subsection{Electrical measurements}
All electrical measurements were performed on a MPITS2000-SE probe station using a Keithley 4200A-SCS parameter analyzer equipped with current preamplifiers for quasi-static measurements and DC biasing. Prior to testing the oscillators, the VO$_2$ nanosheets were preconditioned using a current sweep to redistribute any defects that may have been introduced during the IBE process and to ensure proper IMT behavior. An external arbitrary waveform generator (HP 33120A) was used to drive the JLFET gate in dynamic mode. Time-domain voltage waveforms were acquired using a mixed-signal oscilloscope (Tektronix MSO64), with the acquisition time triggered on $V_{SS}$. 

\subsection{Finite element modeling of VO$_2$ memristors}
Generic two-terminal devices based on the fabricated geometry were simulated in COMSOL Multiphysics to explore the scalability of the system in terms of power consumption. The 3D model represents the nanosheet and the residual layer, placed on a SiO$_2$ substrate (60~nm thick). The study is stationary, and the physics are Electrostatics (Electric Currents) and Heat Transfer (Heat Transfer in Solids), coupled trough Electromagnetic Heating. VO$_2$ electrical conductivity was modeled based on the values measured at different temperatures for devices (W$_0$ = 2 \textmu m, L$_0$ = 3 \textmu m) in the low-power cluster range, while the thermal conductivity was based on literature data \cite{kizuka2015thconductivity}. A thermal contact boundary condition of resistance R$_{th}$ = 1.5 M$\tfrac{K}{W}$ was placed at the VO$_2$/SiO$_2$ interface, with the bottom SiO$_2$ facet set at T$_0$ = 20 °C. The thermal resistor was estimated based on the values extracted from the required $P_{TH}$ to switch the VO$_2$ state at a given temperature, and it was modeled to account for the heat-resistance spread across the different layers of the substrate stack. Particularly, the VO$_2$/SiO$_2$ interfacial thermal resistance was identified as the dominant contribution to $R_{th}$. To account for the dependence of the threshold voltage ($V_{TH}$) and threshold current ($I_{TH}$) on device geometry, the model adjusts the thermal resistance according to the values of W and L relative to the reference dimensions W$_0$ and L$_0$. $V_{TH}$ and $I_{TH}$ are estimated based on the following procedure: the voltage is applied to the electrodes, and the current and internal temperature of the nanosheet are extracted at every voltage value in the center of the channel. The temperature trend with the voltage is fitted, and the voltage for which the temperature reaches the treshold temperature ($T_{TH}$ = 0.9 $T_{IMT}$) is defined as $V_{TH}$. The corresponding $I(T_{IMT}) = I(V_{TH})$ is then defined as $I_{TH}$. A visual representation of the extraction method is reported in the Supporting Information, together with a short discussion about $V_{TH}$ and $I_{TH}$ approximate values as a function of the normalize $R_{th}$ on the contact area A$_c$ ($R_{th0}$).

\subsection{Simulation of 1T-1MR spiking systems and randomness analysis}

The analysis of the measured waveforms was performed by defining a trigger level and identifying signal edges, allowing segmentation of individual oscillation cycles. From each cycle, the period and frequency were extracted, along with the maximum voltage ($V_{TH}$) and minimum voltage ($V_{HL}$). The energy per cycle was obtained by integrating the product of the voltage signal and the bias current over one period.

The rising and falling phases of each cycle were fitted using a standard capacitor charge/discharge model:
\begin{equation}
    V_C\left(t\right) = \left(V_0 - V_a\right)\exp\left(-\frac{t-t_0}{\tau}\right) + V_a
    \label{Eq: CapaCharge}
\end{equation}
The fitting parameters - namely the initial voltage ($V_0$), the asymptotic voltage ($V_a$), the time constant ($\tau$), and the initial time ($t_0$) - were determined for both phases. Using the estimated capacitance ($C_L$), the corresponding resistances and offset voltages for the insulating (rising) and metallic (falling) states were then calculated as:
\begin{equation}
    R_{i/m, eq} = \tau_{r/f}/C_L \quad V_{oi/om} = V_{ar/af}-IR_{i/m},
\end{equation}
where $I$ is the bias current.
For the simulation of the falling escape time, $10000$ discrete iterations were performed to reconstruct the histograms shown in Fig.~\ref{fig:Figure_5}f-h. In each iteration, Eq.~\ref{Eq: CapaCharge} was used to model the noise-free falling phase of the cycle. The median values of the falling time constant ($\tau_f$) and asymptotic voltage ($V_{af}$) extracted from the corresponding measurements were used as input parameters. The initial voltage ($V_{0f}$) was set to 50 mV higher than $V_{af}$ and the initial time was fixed to $t_0 = 0$~s.

A band-limited $1/f$ noise signal was then generated and superimposed onto the noise-free waveform. For each simulation, the holding voltage was selected following a normal distribution mimicking the one observed in measurements. To account for the observed time dependence of the holding voltage, an equivalent thermal RC delay was introduced, with the asymptotic value given by the holding voltage randomly selected for the iteration. The falling escape time was defined as the time corresponding to the first crossing between the noisy signal and the instantaneous holding voltage.

\section{Associated content}
\subsection{Supporting information}
Supporting Information is available containing details of the design, fabrication, and modeling of VO$_2$ memristors; mobility extraction and temperature dependence of the SOI JLFETs characteristics; and a description of the analytical model and statistical simulation of the analog spiking neurons.

\subsection{Author contribution} 
F.B., V.C., and A.I. developed the nanofabrication process, designed, and fabricated the devices. F.B., C.M., V.C., and A.I performed the electrical measurements and analyzed the data. C.M. derived the analytical model and simulated the oscillation dynamics. V.C. performed FEM simulations of VO$_2$ devices. V.C. and A.V. analyzed the variability and optimized the PLD deposition of VO$_2$ films. N.M. prepared the TEM lamellae and characterized them. E.A. worked on the material deposition and characterization tools. F.B., C.M., and V.C. wrote the manuscript with input from the authors. F.B. and A.M.I. conceived the project. A.V., I.S., and A.M.I. supervised the project. F.B., C.M., and V.C. contributed equally to this work. 
\subsection{Notes}
The authors declare no competing financial interest.

\section*{Acknowledgements}
The authors gratefully acknowledge Prof. Pasquale Scarlino (EPFL) and Prof. Edoardo Charbon (EPFL) for their support with the measurement setup. The authors also thank Edoardo Tenna for fruitful discussions on the fabrication of JLFETs. This work was co-founded by the Swiss SERI via the Swiss Chips project D26116, by the SNSF via projects NEMO Synergy CRSII5-209454 and SNSF 200021-208233, and by ERC Synergy under agreement N.~101119062 - SWIMS.

\newpage
\printbibliography

\end{refsection}

\clearpage
\begin{refsection}
\setcounter{figure}{0}
\setcounter{table}{0}
\setcounter{equation}{0}
\renewcommand{\thefigure}{S\arabic{figure}}
\renewcommand{\thetable}{S\arabic{table}}
\renewcommand{\theequation}{S\arabic{equation}}

\begin{center}
{\LARGE\ Supporting Information\\[0.5em]
Monolithically Integrated VO$_2$ Mott Oscillators for Energy-Efficient Spiking Neurons\par}
\vspace{1.5em}
{\large Fabio Bersano$^{1\dagger}$, Cyrille Masserey$^{1\dagger}$,
Vanessa Conti$^{1\dagger}$, Andrea Iaconeta$^1$, Niccolò Martinolli$^1$,
Ehsan Ansari$^1$, Anna Varini$^1$, Igor Stolichnov$^1$,
Adrian Mihai Ionescu$^1$\par}
\vspace{1em}
{$^1$Institute of Electrical and Micro Engineering,
École Polytechnique Fédérale de Lausanne (EPFL), Lausanne, 1015, Switzerland\par}
\vspace{1em}
{$\dagger$ These authors contributed equally to this work.\par}
\end{center}
\vspace{1.5em}

\subsection{Design, fabrication and modeling of VO$_2$ memristors}
\begin{figure}[ht]
    \centering
    \includegraphics[width=0.6\linewidth]{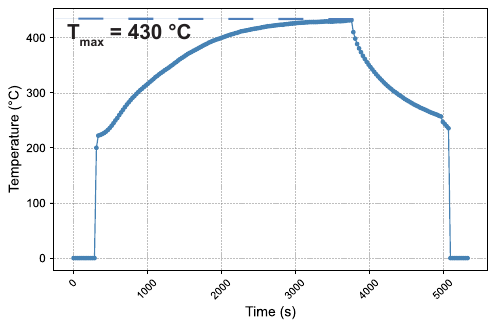}
    \caption{\textbf{PLD thermal budget.} Temperature profile measured during PLD deposition (ramp up, deposition, and ramp down). The baseline temperature is set to 400~$^\circ$C and the maximum temperature approaches 430~$^\circ$C.}
    \label{fig:Tprofile}
\end{figure}
\begin{figure}[ht]
    \centering
    \includegraphics[width=\linewidth]{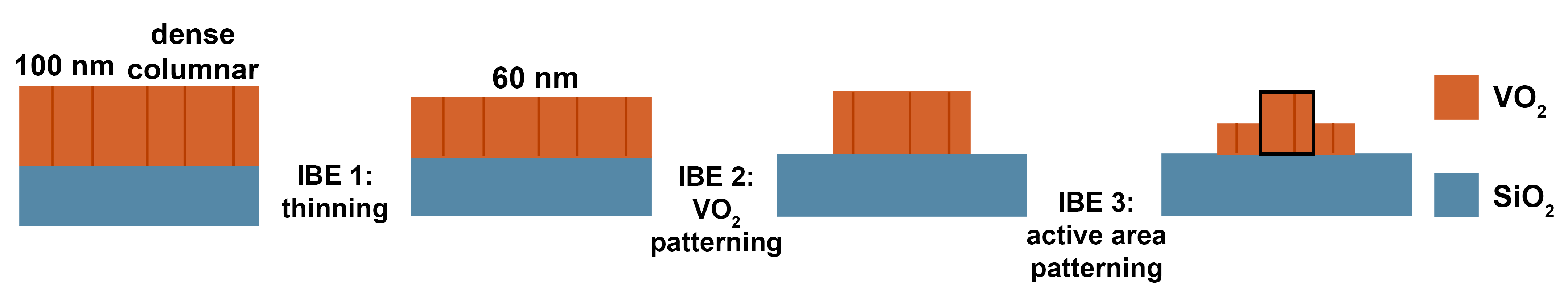}
    \caption{\textbf{Etching of VO$_2$ thin films.} Schematic illustration of the three IBE steps used for patterning VO$_2$ devices on top of SOI JLFETs.}
    \label{fig:IBE_FLOW}
\end{figure}

\begin{figure}[ht]
    \centering
    \includegraphics[width=0.8\linewidth]{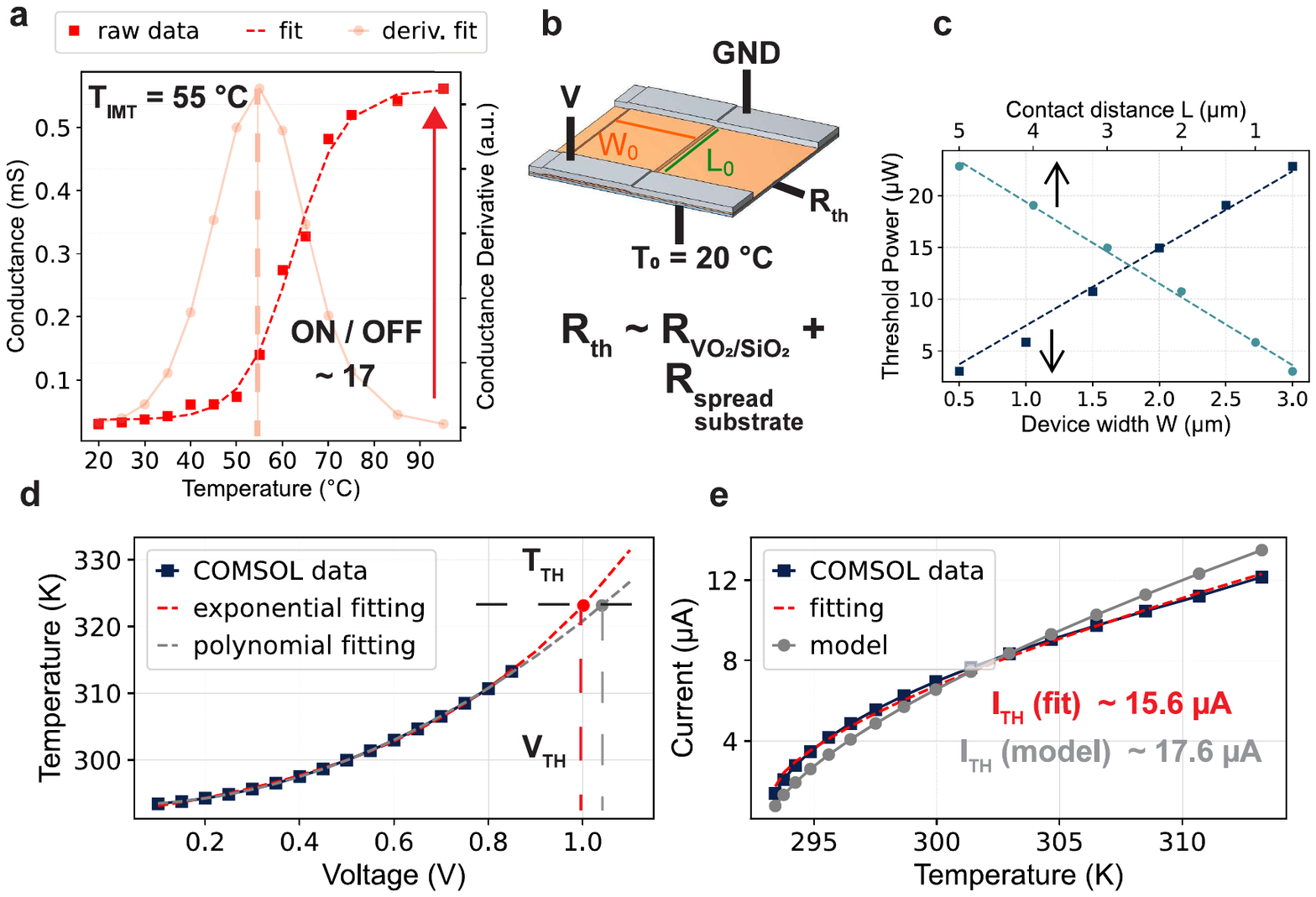}
    \caption{\textbf{Thermal characteristics of device behavior and FEM analysis.} \textbf{a} Representative conductance-temperature curve (forward sweep) of a measured device. Raw data, fitted curve, and its derivative are shown, along with the extracted T$_{IMT}$ and on/off ratio. \textbf{b} Three-dimensional finite-element model of a VO$_2$ nanosheet with nominal dimensions W$_0$ and L$_0$, including boundary conditions detailed in Methods. \textbf{c} Extracted P$_{TH}$ values from COMSOL Multiphysics simulations (W$_0$ = 2~\textmu m, L$_0$ = 3~\textmu m), obtained by varying either W or L while keeping the other parameter fixed. The VO$_2$ conductivity is derived from the fitted conductance curves in (a). \textbf{d} Representative internal temperature-voltage curve obtained from FEM simulations. V$_{TH}$ is determined by fitting the curve at the point where T reaches T$_{TH}$. \textbf{e} Bias current as a function of internal device temperature. The extracted threshold current I$_{TH}$ = I(T$_{TH}$), obtained from fitting and the analytical model, is shown.}
    \label{fig:COMSOL}
\end{figure}
Fig.~\ref{fig:Tprofile} shows the temperature profile extracted from the log-file of the PLD machine, with a peak temperature not exceeding 430~$^\circ$C. The sequence of the IBE steps to fabricate the VO$_2$ two-terminal devices is reported in Fig.~\ref{fig:IBE_FLOW}: after deposition the film is thinned from 100 nm to 60 nm (IBE 1), then VO$_2$ is patterned to leave a 60~nm thick layer with a rectangular footprint, $A_c = W_cL_c$ (IBE 2). In the final step, the active area (device channel) is defined by further thinning the surrounding VO$_2$.\newline

Fig.~\ref{fig:COMSOL}a report an example of thermal characteristic of a VO$_2$ nanosheet. The curve was extracted by injecting a small current (50~nA) and measuring the potential drop at each temperature. The raw data are fitted using the following model:
\begin{equation}
\begin{aligned}
G(T) = G_i + (G_m - G_i)\times f(T) \\
f(T) = \frac{1}{1+\exp{\frac{T_{IMT}-T}{\Delta T}}}
\end{aligned}
\label{GT_eq}
\end{equation}
With G$_m$ and G$_i$ being the the metallic and insulating conductance and $\Delta T$ being the temperature range for which the transition from G$_i$ to G$_m$ occurs. The fitting is used to model VO$_2$ conductivity in COMSOL Multiphysics to reproduce a typical nanosheet device (Fig.~\ref{fig:COMSOL}b) and evaluating the effect of geometry scaling on P$_{TH}$ (Fig.~\ref{fig:COMSOL}c). As explained in Methods section, V$_{TH}$ and I$_{TH}$ can be estimated by extracting V and I values for a temperature equal to T$_{TH}$. An example is reported in Fig.~\ref{fig:COMSOL}d for V$_{TH}$. Fig.~\ref{fig:COMSOL}e shows the current trend with the internal temperature of the device as computed from the finite-element model, its fitting, and the equivalent current estimated from the following analytical model:
\begin{equation}
\begin{aligned}
I(T) \approx\frac{1}{2}\sqrt{\frac{3R_R}{R_R-1}}\sqrt{\frac{1}{\rho_{20^\circ C}}}\sqrt{\frac{A_c}{R_{th0}}}\sqrt{\frac{A_{CS_{dev}}}{L_{dev}}}(T_{TH}-T)^{\alpha=2/3}\\
\end{aligned}
\label{IT_model}
\end{equation}

Eq.~\ref{IT_model} is derived from the more common expression for I$_{TH}$ in the Joule heating framework, where the current scales with $\alpha = \frac{1}{2}$~\cite{conti2026electroforming}. The model offers an alternative closer to the computational data proposed by the FEM simulation, and it relies on geometrical parameters, such as the cross-section area $A_{CS_{dev}}$ or the length of the device $L_{dev}$, the transition magnitude R$_R$, the specific thermal resistance $R_{th0}$ and the easily accessible value of the room temperature resistivity $\rho_{20^\circ C}$. The threshold current can be computed as:

\begin{equation}
I_{TH} = \sqrt{\frac{G_{th}(T_{TH}-T)}{R_{TH}}} = \sqrt{\frac{(g_{eff}t_{eff})(T_{TH}-T)}{\rho_{TH}}\frac{L_{dev}}{A_{CS_{dev}}}} \approx c\times\sqrt{\frac{t_{eff}A_{CS_{dev}}}{\rho_{20 ^\circ C}L_{dev}}}(T_{TH}-T)^{2/3}
\label{eq:i}
\end{equation}
\begin{equation}
G_{th} = g_{eff}t_{eff} = \frac{A_c}{R_{th0}}
\label{eq:ii}
\end{equation}

\begin{equation}
c = \sqrt{\frac{\rho_{20 ^\circ C}}{\rho_{TH}}g_{eff}}(T_{TH}-T)^{-1/6} \approx \sqrt{\frac{3R_R}{R_R-1}}\sqrt{g_{eff}}\frac{1}{2},
\label{eq:iii}
\end{equation}

where Eq.~\ref{IT_model} can be derived by substituting in Eq.~\ref{eq:i} the expressions in Eq.~\ref{eq:ii} for the thermal conductivity and Eq.~\ref{eq:iii} for the equivalent $c$ term. We underline that the approximation in Eq.~\ref{eq:iii} is based on the following assumptions: 1) considering T close to room temperature, the temperature depending term converges to value close to $\frac{1}{2}$ for T$_{TH}$ in a wide range between 55 $^\circ$C and 75 $^\circ$C, which is typical for VO$_2$-based devices. 2) $\rho_{TH}$ is the resistivity at the transition point, which can be expressed as the semi-difference between the resistivity when the transition starts $\rho_{s}$ and ends $\rho_{e}$ in the $\Delta T$ range. $\rho_{e}$ can be expressed as $\rho_{m}$, considering the limited change of resistivity observed after the transition. At the same time, $\rho_{20 ^\circ C}$ can be expressed in function of $\rho_{s}$:
\begin{equation}
\rho_{TH} = \frac{\rho_s - \rho_m}{2}=\rho_s\frac{R_R-1}{2R_R}
\label{eq:i2}
\end{equation}
\\
\begin{equation}
\frac{\rho_{20^\circ C}}{\rho_{TH}}=\frac{(\rho_s+\Delta \rho)(2R_R)}{\rho_s(R_R-1)}=\frac{(1+\frac{\Delta \rho}{\rho_s})(2R_R)}{R_R-1}\approx\frac{(1+\frac{1}{2})(2R_R)}{R_R-1}
\label{eq:ii2}
\end{equation}
Where the latter approximation in Eq.~\ref{eq:ii2} is based on experimental values observed in our films. Based on the provided equations, it is possible to find an equivalent expression for the threshold power:
\begin{equation}
    P_{TH} \approx \frac{A_c}{2R_{th0}}(\frac{R_R}{R_R-1})(T_{TH}-T)^{4/3}
\end{equation}
Furthermore, following the definition provided in~\cite{conti2026electroforming} for the spike temperature (T$_{spk}$) at the IMT point, it is possible to compute the ratio between the spike temperature generated for the P$_{TH}$ needed to switch the state of the device:
\begin{equation}
    \frac{T_{spk}}{P_{TH}} \approx 2\frac{R_{th0}}{A_c}(R_R-1)(T_{TH}-T)^{-1/3}
\end{equation}
As a consequence, despite a larger A$_c$ results into an increased power consumption, the device benefits from a lower temperature spike, thus higher stability in-operation. In our nanosheet, the larger $A_c$ is achieved through the fabrication a the residual layer, whose reduced thickness is designed to minimally contribute to the overall power consumption. 

\newpage
\subsection{Electrical characterization of SOI JLFETs and temperature dependence}

\begin{figure}[h]
    \centering
    \includegraphics[width=0.95\linewidth]{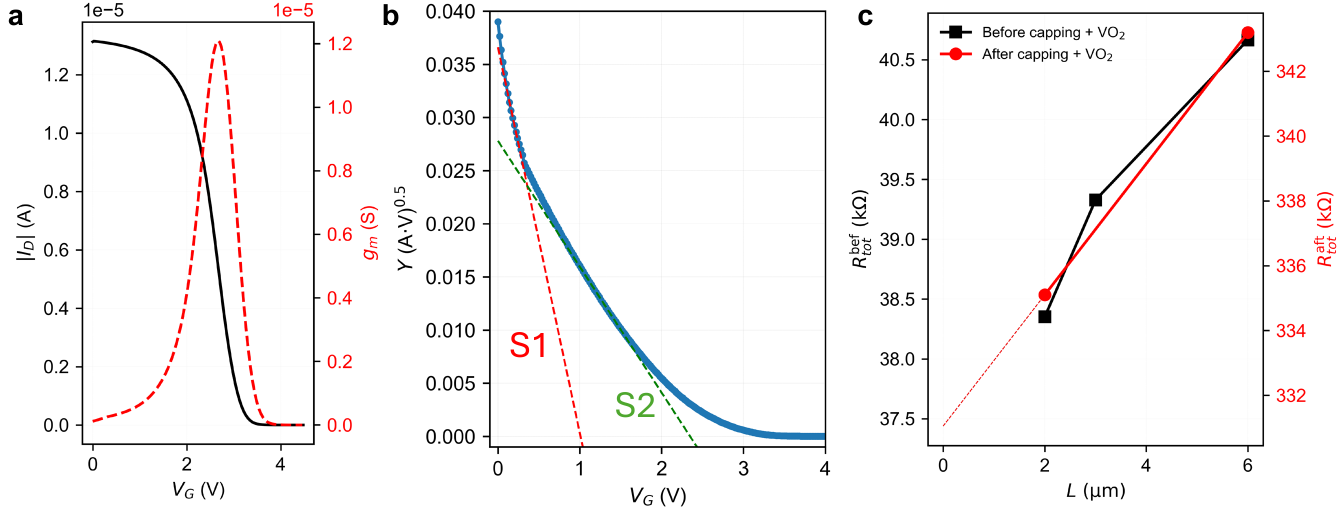}
    \caption{\textbf{Electrical parameter analysis of the fabricated JLFETs.} \textbf{a} Transfer characteristic and transconductance measured at V$_{DS} = - 0.5~V$. \textbf{b} Computed Y-function showing two distinct slopes associated with the accumulation regime (S1) and channel conduction (S2). \textbf{c} Extracted total series resistance of the fabricated transistors.}
    \label{fig:SI_JLFETs_RSD}
\end{figure}

Fig.~\ref{fig:SI_JLFETs_RSD}a shows an example of a measured transfer curve, along with the extracted transconductance $g_m$ as a function of the applied front-gate voltage. The Y-function was computed as $ Y=\frac{I_D}{gm^{1/2}}$. As thoroughly discussed in \cite{jeon2013revisited}, two distinct slopes can be identified in the $Y(V_G)$ characteristics of junctionless FETs, showing evidence of accumulation regime and bulk conduction, differently from inversion-mode transistors (S1 and S2 in Fig.~\ref{fig:SI_JLFETs_RSD}). The effective hole mobility $\mu_{eff}$ was extracted from the slope at lower gate voltages (corresponding to the accumulation regime - S1 in Fig.~\ref{fig:SI_JLFETs_RSD}b) using the following relationship:
\begin{equation}
\Big(\frac{dY}{dV_G}\Big)^2 = \frac{W}{L}\cdot C_{ox} \cdot \mu_{eff} \cdot V_{DS},   
\end{equation}
where $C_{ox}$ is the gate oxide capacitance. 
The extracted total series resistance of the fabricated transistors is reported in Fig.~\ref{fig:SI_JLFETs_RSD}c.

\begin{figure}[ht]
    \centering
    \includegraphics[width=0.95\linewidth]{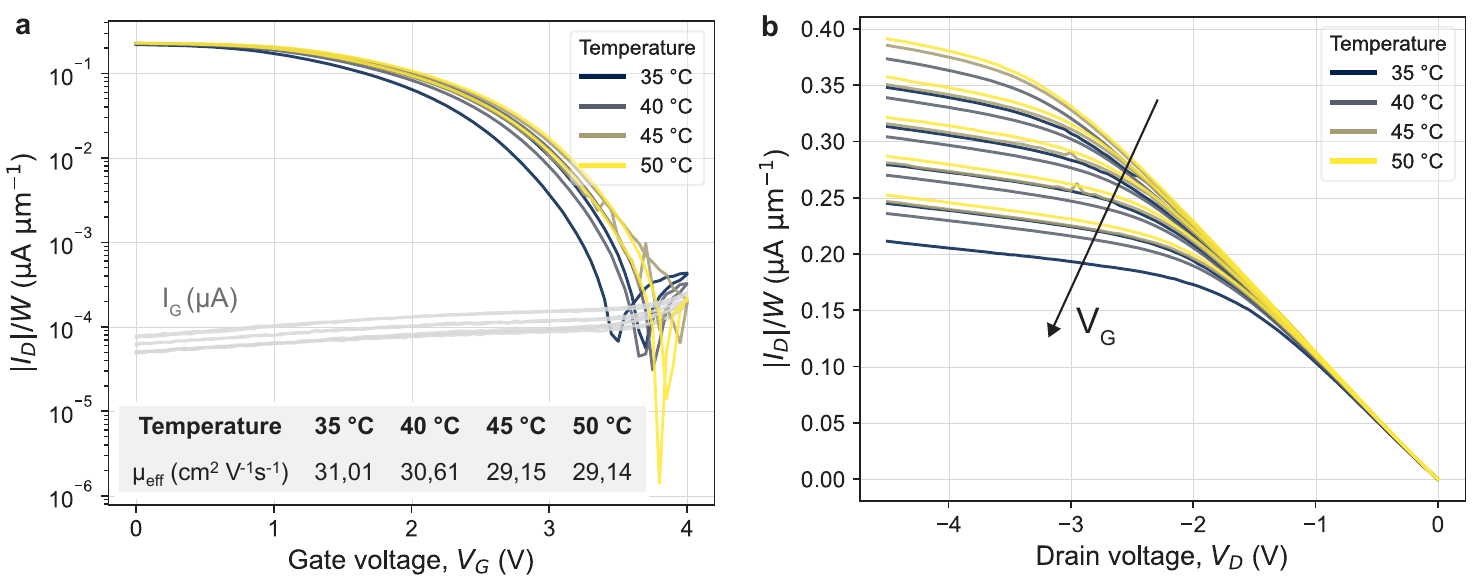}
    \caption{\textbf{SOI JL-FETs characterized at different temperatures.} \textbf{a} Transfer characteristics measured at V$_{DS}$ = $-1$~V, together with the gate leakage current (I$_G$). The inset shows the extracted effective hole mobility obtained using a modified Y-method~\cite{jeon2013revisited}. \textbf{b} Output characteristics measured for V$_G$ ranging from 0 to 1~V in steps of 0.25~V. The source contact is to 0~V. }
    \label{fig:SI_JLFETs}
\end{figure}

Fig.~\ref{fig:SI_JLFETs} shows the electrical characterization of a JLFET transistor (with width W=30~\textmu m and length L=2~\textmu m). The saturation current increases with temperature, as expected from the increased carrier concentration and enhanced thermionic emission. While temperature affects the subthreshold slope, the hole mobility remains nearly unchanged, consistent with bulk conduction in the highly doped SOI substrate.

\newpage
\subsection{Analytical model and statistical simulation of analog spiking neurons}

\begin{figure}[ht]
    \centering
    \includegraphics[width=0.8\linewidth]{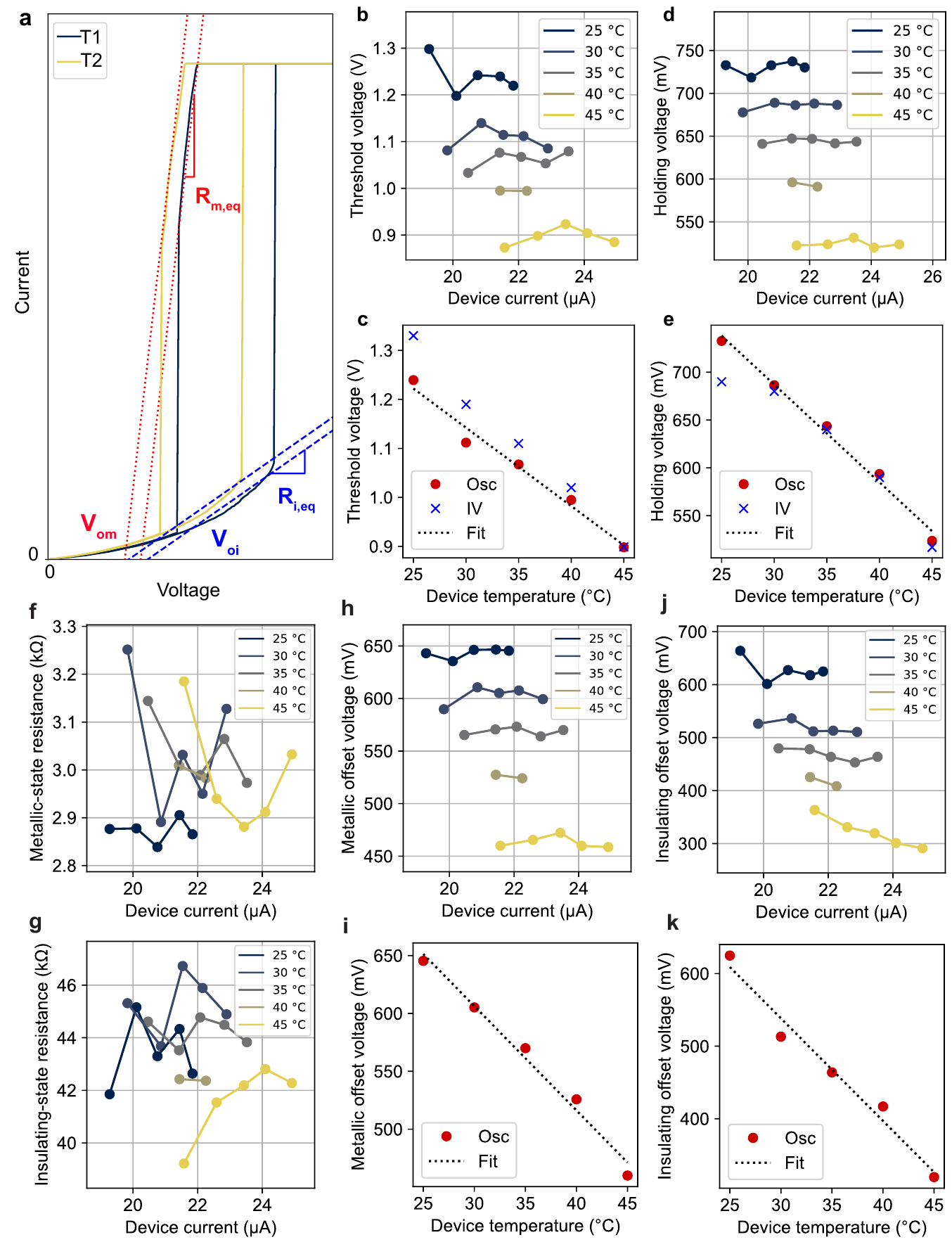}
    \caption{\textbf{Model description and supplementary simulation results.}
    \textbf{a} Schematic illustration of the analytical model (for illustrative purposes only, with $T_1>T_2$).
    \textbf{b-c} Current and temperature dependence of the threshold voltage.
    \textbf{d-e} Current and temperature dependence of the holding voltage. In (c) and (e), the red dots correspond to values extracted from the oscillation waveforms, while the blue crosses represent the average of 30 quasi-static I-V measurements. The dotted line shows the linear fit used for the analytical model. 
    \textbf{f-g} Current and temperature dependence of the  equivalent metallic and insulating resistances.
    \textbf{h-i} Current and temperature dependence of the metallic offset voltage.
    \textbf{j-k} Current and temperature dependence of the insulating offset voltage.
    }
    \label{fig:SI_Model}
\end{figure}

\textbf{Insulating- and metallic-state resistance}\\
Fig.~\ref{fig:SI_Model} illustrates the model described in the main text, together with the current and temperature dependence for all six parameters included in the analytical model. In Fig.~\ref{fig:SI_Model}a, the blue and red lines highlight the affine functions used to approximate the nonlinear I-V characteristics. The equivalent insulating-state resistance, $R_{i,eq}$ represents an approximation of the slope between the hold and threshold voltages, while $V_{oi}$ corresponds to the offset voltage capturing the nonlinearity (in blue). A similar approximation is used for the metallic state (in red). In the illustrative example, the two curves at different temperatures exhibit similar slopes in both the insulating and metallic states, consistent with the measurements in Fig.~2f. As detailed in Methods, the model parameters were extracted by fitting the oscillation waveforms instead of the quasi-static I-V curves, since the latter do not capture
transient electrothermal effects such as self-heating and heat dissipation. This is evidenced by the different values of V$_{TH}$ and V$_{HL}$ extracted
from dynamic and quasi-static characterization in Fig.~\ref{fig:SI_Model}c,e. 

\begin{figure}[ht!]
    \centering
    \includegraphics[scale=0.7]{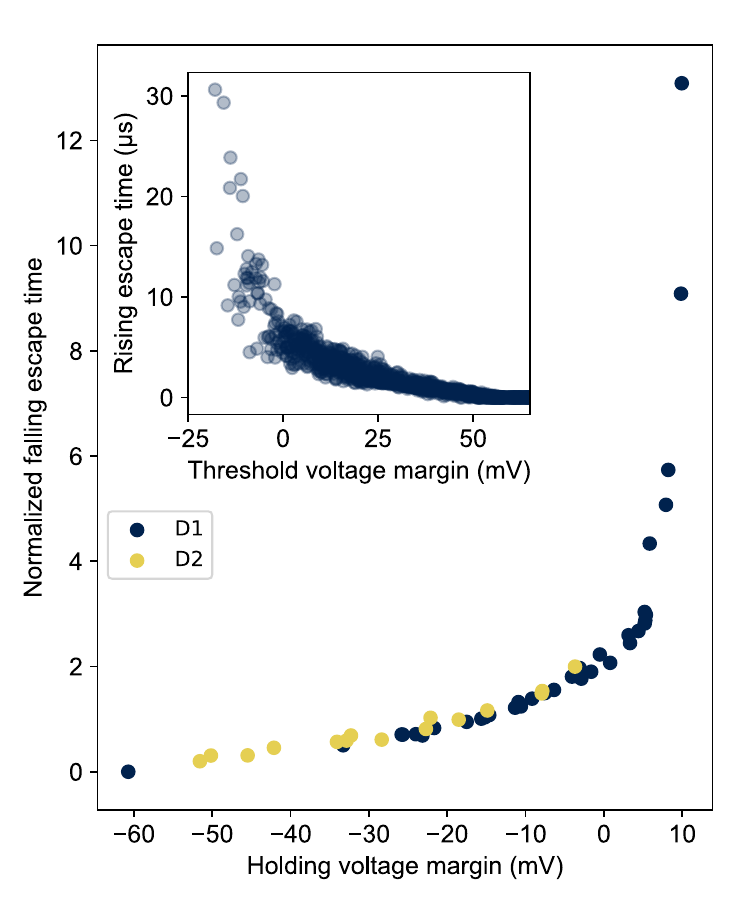}
    \caption{\textbf{Generalization of escape-time behavior.} The falling escape time, normalized by the time constant $\tau$, is shown for two different devices (D1 corresponds to the device presented in Fig.~5). Each point represents the median over all measured cycles. Inset: dynamic characterization of the rising escape time as a function of the threshold voltage margin under sinusoidal gate-voltage excitation (see Fig.~6d) (each point corresponds to one oscillation cycle).}
    \label{fig:SI_IncT}
\end{figure}

\textbf{Escape time}\\
The escape-time behavior shown in Fig.~\ref{fig:SI_IncT} is observed across all measured devices, for both biasing near the threshold and holding voltages. As illustrated in the inset of Fig.~\ref{fig:SI_IncT}, similar trends are obtained under dynamic excitation conditions, confirming the generality of this behavior. Direct normalization of the escape-time characteristics is nontrivial, as different regions of the curve depend on both the intrinsic time constant $\tau$ of the device and the noise level of the measurement setup. In Fig.~\ref{fig:SI_IncT}, normalization is performed using the extracted $\tau$ for each device.
\begin{figure}[ht!]
    \centering
    \includegraphics[width=\linewidth]{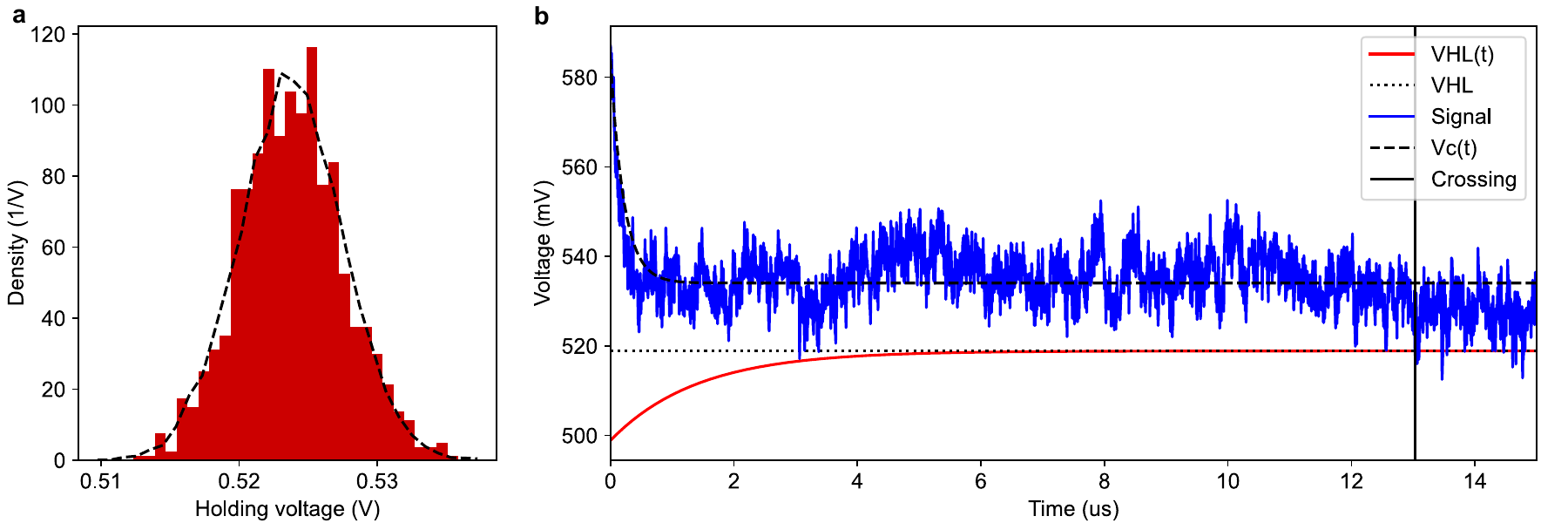}
    \caption{\textbf{Statistical simulation illustration. a} Holding voltage distribution extracted from measurements (red) and the corresponding distribution from the simulations (black dashed line). \textbf{b} Example of a single iteration of the statistical simulation demonstrating noise-induced triggering.}
    \label{fig:SI_StatisitcSimulation}
\end{figure}

\textbf{Noise and time-dependent holding voltage}\\
Fig.~\ref{fig:SI_StatisitcSimulation} illustrates one iteration of the stochastic spiking simulation. In Fig.~\ref{fig:SI_StatisitcSimulation}b, the selected holding voltage for this iteration (V$_{HL}$) is shown as a black dotted line, while the time-dependent hold voltage (V$_{HL(t)}$) is shown in red. The noise-free analytical electrical signal ($VC(t)$) (corresponding to Eq.~3 is plotted as a black dashed line, and the simulated signal including limited bandwidth $1/f$ noise (Signal) is shown in blue. The vertical black line indicates the first crossing event between the signal and the holding voltage (Crossing), which would trigger the next oscillation cycle. This example highlights the necessity of introducing a time-dependent holding voltage. Without this effect, the crossing event, and thus the triggering of the next cycle, would occur significantly earlier than observed experimentally. We assume that the time dependence of the holding voltage arises from thermal relaxation processes associated with Joule heating during switching. Accordingly, a thermal RC model is adopted as a first-order approximation.

\newpage
\printbibliography

\end{refsection}
\end{document}